\documentclass[twoside, a4paper, notoc]{JHEP3}
\usepackage{amsfonts}
\usepackage{amssymb}
\usepackage{comment}
\usepackage{epsfig}
\usepackage{graphicx}
\usepackage{amsmath}

\newcommand{\be}{\begin{equation}}
\newcommand{\ee}{\end{equation}}
\newcommand{\bea}{\begin{eqnarray}}
\newcommand{\eea}{\end{eqnarray}}
\newcommand{\mbb}{\mathbb}
\newcommand{\ti}{\times}
\newcommand{\half}{\frac{1}{2}}
\newcommand{\mc}{\mathcal}

\newcommand{\beqa}{\begin{eqnarray}}
\newcommand{\eeqa}{\end{eqnarray}}

\title{Systematics of String Loop Corrections in Type IIB Calabi-Yau Flux
Compactifications}
\author{Michele Cicoli, Joseph P. Conlon and Fernando Quevedo \\
DAMTP, Centre for Mathematical Sciences, \\
Wilberforce Road, Cambridge, CB3 0WA, UK. \\
Email: \email{M.Cicoli@damtp.cam.ac.uk,
J.P.Conlon@damtp.cam.ac.uk, F.Quevedo@damtp.cam.ac.uk}}

\abstract{ We study the behaviour of the string loop corrections
to the $N=1$ $4D$ supergravity K\"{a}hler potential that occur in
flux compactifications of IIB string theory on general Calabi-Yau
three-folds. We give a low energy interpretation for the
conjecture of Berg, Haack and Pajer for the form of the loop
corrections to the K\"{a}hler potential. We check the consistency
of this interpretation in several examples. We show that for
arbitrary Calabi-Yaus, the leading contribution of these
corrections to the scalar potential is always vanishing, giving an
``extended no-scale structure''. This result holds as long as the
corrections are homogeneous functions of degree $-2$ in the
2-cycle volumes. We use the Coleman-Weinberg potential to motivate
this cancellation from the viewpoint of low-energy field theory.
Finally we give a simple formula for the 1-loop correction to the scalar potential in terms
of the tree-level K\"ahler metric and the conjectured correction to the K\"ahler potential.
We illustrate our ideas with several examples. A companion paper will use these
results in the study of K\"{a}hler moduli stabilisation.}

\preprint{DAMTP-2007-75}

\begin{document}

\tableofcontents

\bigskip

\section{Introduction}

Four-dimensional effective actions have played a major r\^{o}le in
addressing the moduli stabilisation problem of string
compactifications (for recent reviews with many references see
\cite{dk,  germans}).
 Most of these efforts rely on Calabi-Yau backgrounds with
fluxes of RR fields for which a worldsheet understanding of string
interactions is not available and the effective action approach is
the only  reliable tool at present to study the moduli dynamics.
 For moduli stabilisation and the study of other low-energy
and cosmological implications,
 it is then important to have control on the ${\cal N} =1$
supergravity effective actions associated to string
compactifications.

The K\"{a}hler potential $K$ is the least understood part of these
four-dimensional effective actions. There has been substantial
progress in determining the tree-level structure of the K\"{a}hler
potential as a function of the many moduli fields appearing in
arbitrary Calabi-Yau compactifications \cite{louis}. However,
unlike the superpotential $W$, the lack of holomorphy implies that
the K\"{a}hler potential can receive corrections to all orders in
the $\alpha'$ and $g_{s}$ expansions. The presence of no-scale
structure in the K\"{a}hler potential makes understanding the
K\"{a}hler corrections particularly pressing, as it is the
corrections that give rise to the leading perturbative terms in
the scalar potential.

Mirror symmetry and the underlying ${\cal{N}}=2$ structure was
used to extract the leading order $\alpha'$ corrections
\cite{bbhl}. Explicit string amplitude calculations to determine
the  loop corrections to $K$ are not available for general fluxed
Calabi-Yau compactifications, and only simple unfluxed toroidal
orientifold cases have been used for concrete computations
\cite{bhk}.

Despite the difficulty of explicitly computing loop corrections in general
Calabi-Yau flux backgrounds, given their importance it is necessary to try and
go as far as possible. In this respect we observe that there is an easier and a harder
part to computing the form of loop corrections. The easier part involves the parametric scaling
of moduli that control the loop expansion - in IIB these are the dilaton, which controls the string coupling,
and the K\"ahler moduli, which controls the gauge coupling on D7 branes. The harder part
involves the actual coefficients of the loop expansion, which depend on the complex structure moduli and would require
a explicit string computation. This article focuses entirely on the `easier' part; however
as the K\"ahler moduli are unstabilised at tree-level, such knowledge is very important for moduli stabilisation.

Recently, Berg, Haack and Pajer (BHP) \cite{bhp} gave arguments
for the general functional dependence of the leading order loop
corrections to $K$ on the K\"{a}hler moduli. By comparing with the
toroidal orientifold calculations and the standard transformations
required to go
 from the string frame, where string amplitudes are computed, to the
 physical
 Einstein frame that enters the supergravity action, they conjectured the
parametric form of the leading corrections for general Calabi-Yau
compactifications as a function of the K\"{a}hler moduli. As mentioned above, it is
this dependence (on the K\"{a}hler moduli) that is more relevant
for moduli stabilisation, as the dilaton and complex structure
moduli are usually stabilised directly from the fluxes and it is
only the K\"{a}hler moduli that need quantum corrections to the
scalar potential to be stabilised. These quantum corrections
include non-perturbative corrections to the superpotential $W$
(since $W$ is not renormalised perturbatively \cite{beq}),
$\alpha'$ and string loop corrections to $K$. Non-perturbative
corrections to $K$ are subdominant with respect to the
perturbative corrections. It is then of prime importance to have
control on the parametric form of the quantum corrections to $K$.

In this article we study in detail the leading order loop
corrections to $K$ conjectured by BHP. We provide a low-energy
interpretation of this conjecture and give a general argument that, with the conjectured form,
the leading loop corrections to the K\"{a}hler potential cancel at
leading order in their contributions to the scalar potential. This
is very relevant for the robustness of the large volume scenario
\cite{bbcq} for which the leading $\alpha'$ corrections were used
to obtain stabilised exponentially large volumes. Even though the
leading string loop correction to $K$ is dominant over the
$\alpha'$ corrections, its contribution to the scalar potential is
subdominant \cite{bhk, bhp}. We will also see that this cancellation
is necessary if the loop corrections to $K$ are to generate corrections to the scalar potential
consistent with those expected from the
Coleman-Weinberg potential. We also extend this result to more
general possible corrections to $K$, showing that the only
property needed for the cancellation is that $\delta K$ is a homogeneous
function of degree $n=-2$, which includes the BHP proposal. We
illustrate our results with several examples.

This article is organized as follows. In Section \ref{sec2} we
review the present status of the tree-level and quantum corrected
effective actions and their r\^{o}le for moduli stabilisation.
Sections \ref{sec3} and \ref{sec4} are the main parts of the
article in which we study in detail the proposed form of the string loop corrections to
the K\"{a}hler potential, their interpretation in terms of the
Coleman-Weinberg potential and examples of different Calabi-Yau
manifolds where these corrections are relevant. Finally in a
comprehensive appendix \ref{appendix} we provide a general
discussion of the different proposals that have been put forward
to stabilise K\"{a}hler moduli, emphasizing that in all cases it
is necessary to understand the quantum corrections to $K$. This is
an explicit illustration of the importance to better understand
the perturbative corrections to the supersymmetric action. In
particular the `extended no-scale structure' of Section \ref{sec4}
is crucial to establish the robustness of the exponentially large
volume scenario. In a companion article \cite{ccq2} we will use
our results to study moduli stabilisation in different classes of
Calabi-Yau manifolds.

\section{Effective Action for Type IIB Flux Compactifications}
\label{sec2}

\subsection{Tree-level Action}
\label{sec2.1}

We first review the low energy theory of IIB Flux
Compactifications on a Calabi-Yau $X$ \cite{gkp}. The tree-level
superpotential is generated by turning on fluxes and takes the
Gukov-Vafa-Witten form:
\begin{equation}
W_{tree}(S, U)=\int\limits_{X}G_{3}\wedge \Omega ,  \label{Wtree}
\end{equation}%
with $ G_{3}=F_{3}+iSH_{3}$, where $F$ and $H$ are RR and NSNS
3-form fluxes respectively, $S$ is the axio-dilaton $
S=e^{-\varphi }+iC_{0}, $
(with $e^{\varphi }$ the string coupling and $C_{0}$ the RR 0-form), and $%
\Omega $ is the holomorphic (3,0)-form which depends on the
complex structure moduli $U$. The tree level K\"{a}hler potential
$K_{tree}$ is
\begin{equation}
K_{tree}=-2\ln \left( \mathcal{V}\right)-\ln \left(
S+\bar{S}\right) -\ln \left( -i\int\limits_{X}\Omega \wedge
\bar{\Omega}\right) ,  \label{eq}
\end{equation}%
where $\mathcal{V}$ is the Einstein frame internal volume that
depends only on $(T+ \bar{T})$. $K_{tree}$ has a factorized form
with respect to $T$, $U$ and $S$ moduli. The $T$ moduli are
defined by
\begin{equation}
T_{i}=\tau _{i}+ib_{i},  \label{axion}
\end{equation}%
where $\tau _{i}$ is the Einstein frame volume of a 4-cycle
$\Sigma _{i}$, measured in units of $l_{s}=(2\pi )\sqrt{\alpha ^{'
}}$, and $b_{i}$ is the component of the RR 4-form $C_{4}$ along
this cycle:$\int\limits_{\Sigma _{i}}C_{4}=b_{i}$. The 4-cycle
volumes $\tau _{i}$ may be related to the 2-cycle volumes $t_{i}$.
Letting $D_{i}$ be a basis of divisors on $X$ (we use $D_i$ to
denote both the divisor and its dual 2-form), and $k_{ijk}$ the
divisor triple intersections, the overall volume $\mathcal{V}$ can
be written as
\begin{equation}
\mathcal{V}=\frac{1}{6}\int\limits_{X}J\wedge J\wedge J=\frac{1}{6}%
k_{ijk}t^{i}t^{j}t^{k},
\end{equation}
where $J=t^{i}D_{i}$ is the K\"{a}hler form. The 4-cycle volumes
$\tau _{i}$ are defined as
\begin{equation}
\tau _{i}=\frac{\partial \mathcal{V}}{\partial t^{i}}=\frac{1}{2}%
\int\limits_{X}D_{i}\wedge J\wedge J=\frac{1}{2}k_{ijk}t^{j}t^{k}.
\end{equation}%
Finally, let us introduce the following notation%
\begin{equation}
A_{ij}=\frac{\partial \tau _{i}}{\partial
t^{j}}=\int\limits_{X}D_{i}\wedge D_{j}\wedge J=k_{ijk}t^{k}.
\label{aa}
\end{equation}%
Some useful relations that we will use subsequently are
\begin{equation}
\left\{
\begin{array}{c}
t^{i}\tau _{i}=3\mathcal{V}, \\
A_{ij}t^{j}=2\tau _{i}, \\
A_{ij}t^{i}t^{j}=6\mathcal{V},%
\end{array}%
\right.   \label{useful}
\end{equation}%
along with%
\begin{equation}
K_{i}^{0}\equiv \frac{\partial \left( K_{0}\right) }{\partial \tau _{i}}=-%
\frac{t_{i}}{\mathcal{V}},
\end{equation}%
where $K_{0}=-2\ln \left( \mathcal{V}\right) $. In addition, the
general
form of the K\"{a}hler matrix is%
\begin{equation}
K_{ij}^{0}\equiv \frac{\partial ^{2}\left( K_{0}\right) }{\partial
\tau
_{i}\partial \tau _{j}}=\frac{1}{2}\frac{t_{i}t_{j}}{\mathcal{V}^{2}}-\frac{%
A^{ij}}{\mathcal{V}},
\end{equation}%
and its inverse looks like%
\begin{equation}
K_{0}^{ij}\equiv \left( \frac{\partial ^{2}\left( K_{0}\right)
}{\partial
\tau _{i}\partial \tau _{j}}\right) ^{-1}=\tau _{i}\tau _{j}-\mathcal{V}%
A_{ij}.
\end{equation}
For later convenience, we have expressed the derivatives of the
K\"{a}hler potential in terms of derivatives with respect to $\tau
= \hbox{Re}(T)$ rather than derivatives with respect to $T$ (this
accounts for some differences in factors of $2$ in certain
equations compared to the literature).
From the previous relations it is also possible to show that%
\begin{equation}
K_{0}^{ij}K_{i}^{0}=-\tau _{j},  \label{p1}
\end{equation}%
and the more important result%
\begin{equation}
K_{0}^{ij}K_{i}^{0}K_{j}^{0}=3.  \label{no scale}
\end{equation}%
The $\mc{N}=1$ F-term supergravity scalar potential is given by:
\begin{equation}\label{scalarpotential}
V=e^{K}\left\{ K^{SS}D_{S}WD_{S}\bar{W}+K^{UU}D_{U}WD_{U}\bar{W}%
+4 K^{ij}D_{i}WD_{j}\bar{W}-3\left\vert W\right\vert ^{2}\right\}
,
\end{equation}%
where
\begin{equation}
\left\{
\begin{array}{l}
D_{i}W=\frac{\partial W}{\partial \tau _{i}}+\half W\frac{\partial
K}{\partial
\tau _{i}}\equiv W_{i}+\half K_{i}W, \\
D_{j}\bar{W}=\frac{\partial \bar{W}}{\partial \tau _{j}}+ \half \bar{W}\frac{%
\partial K}{\partial \tau _{j}}\equiv \bar{W}_{j}+\half K_{j}\bar{W}.%
\end{array}%
\right.
\end{equation}%
The form of the scalar potential given in (\ref{scalarpotential})
has used the factorization of the
 moduli space: in general this will be lifted by quantum corrections.
As $W_{tree}$ is independent of the K\"{a}hler moduli,
this reduces to%
\begin{equation}
V=e^{K}\left\{
K^{SS}D_{S}WD_{S}\bar{W}+K^{UU}D_{U}WD_{U}\bar{W}+\left(
K^{ij}K_{i}K_{j}-3\right) \left\vert W\right\vert ^{2}\right\}
\label{eqq}.
\end{equation}%
Furthermore, (\ref{no scale}) implies the existence of no scale
structure as the last term of (\ref{eqq}) vanishes:
\begin{equation}
V=e^{K}\left\{
K^{SS}D_{S}WD_{S}\bar{W}+K^{UU}D_{U}WD_{U}\bar{W}\right\} \geq 0.
\end{equation}%
As the scalar potential is positive semi-definite it is
possible to fix the dilaton and the complex structure moduli by demanding $%
D_{S}W=0=D_{U}W$. Usually, these fields are integrated out setting
them equal to their vacuum expectation values but sometimes we
will keep their dependence manifest. However since they are
stabilised at tree level, even though they will couple to quantum
corrections, these will only lead to subleading corrections to
their VEVs, so it is safe just to integrate them out. From now on,
we will set
\begin{equation}
W_{0}=\left\langle \int\limits_{X}G_{3}\wedge \Omega \right\rangle
.
\end{equation}

\subsection{Non-perturbative and $\alpha'$ Corrections}
\label{sec2.2}

As seen in the previous paragraph, at tree level we can stabilise
only the dilaton and the complex structure moduli but not the
K\"{a}hler moduli. The only possibility to get mass for these
scalar fields is thus through quantum corrections.

It is known that in\textit{\ N=1 4D SUGRA}, the K\"{a}hler
potential receives corrections at every order in perturbations
theory, while the superpotential receives non-perturbative
corrections only, due to the non-renormalisation theorem. The
corrections will therefore take the general form:
\begin{equation}
\left\{
\begin{array}{l}
K=K_{tree}+K_{p}+K_{np}, \\
W=W_{tree}+W_{np},
\end{array}
\right.
\end{equation}
and the hope is to stabilise the K\"{a}hler moduli through these
quantum corrections. In this section we will review the behaviour
of the non-perturbative and $\alpha'$ corrections and then study
the $g_{s}$ corrections in the main part of our paper.

Non-perturbative corrections to the superpotential are given by an
infinite series of contributions \be
W_{np}=\sum\limits_{i,m}A_{i,m}(S,U)\, e^{-ma_{i}T_{i}}  . \ee
They can arise from either Euclidean D3-brane instantons
($a_{i}=2\pi $) or gaugino condensation on wrapped D7-branes
($a_{i}=2\pi /N,$ with $N$ the rank of the condensing gauge
group). In general, $A_{i,m}$ depend on both dilaton and complex
structure moduli. We will always work in a regime where $%
a_{i}\tau _{i}\gg 1$ $\forall i=1,...,h_{1,1}$ so that we can
ignore higher
instanton corrections and keep just the leading non-perturbative corrections:%
\begin{equation}
W_{np}=\sum\limits_{i}A_{i}(S,U)e^{-a_{i}T_{i}}. \label{Wnp}
\end{equation}%
$K_{np}$ can come from either worldsheet or brane instantons and
is
subdominant compared to the perturbative corrections to the K\"{a}%
hler potential (see for instance \cite{kaplu,bqq}) which in
general come from both the $\alpha '$ and the $g_{s}$ expansion
\begin{equation}
K_{p}=\delta K_{(\alpha ^{' })}+\delta K_{(g_{s})}.
\end{equation}
The leading $\alpha'$ correction to the K\"{a}hler potential comes
from the ten dimensional $\mathcal{O}(\alpha'^{3})$
$\mathcal{R}^{4}$ term.
It has been computed in \cite{bbhl} and reads%
\begin{eqnarray}
K_{0}+\delta K_{(\alpha')} &=&-2\ln \left( \mathcal{V}+\frac{\xi
}{2}\text{\
}\hbox{Re}\left( S\right) ^{3/2}\right) =  \notag \\
&=&-2\ln \left( \mathcal{V}\right) -\frac{\xi \hbox{Re}\left( S\right) ^{3/2}%
}{\mathcal{V}}+\mathcal{O}\left( 1/\mathcal{V}^{2}\right) ,
\label{eq2}
\end{eqnarray}%
where the constant $\xi$ is given by%
\begin{equation}
\xi =-\frac{\chi (X)\zeta (3)}{(2\pi )^{3}},
\end{equation}%
with $\chi (X)=2\left( h^{1,1}-h^{2,1}\right) $ and $\zeta
(3)\equiv \sum_{k=1}^{\infty }1/k^{3}\simeq 1.2$. We stress the
point that the $\alpha' $ expansion is an expansion in inverse
volume and thus can be controlled only at large volume. This is
important, as very little is known about higher $\alpha' $
corrections, the exact form of which are not known even in the
maximally supersymmetric flat 10D IIB theory. From now on we focus
only on situations in which the volume can be stabilised at
$\mathcal{V}\gg 1$ in order to have theoretical control over the
perturbative expansion in the low-energy effective field theory.
The inclusion of (\ref{Wnp}) and (\ref {eq2}) now gives the
following scalar potential (where the dilaton has been fixed and
the factor $\hbox{Re}\left( S\right) ^{3/2}$ included in the
definition of $\xi$)
\begin{eqnarray}
V &=&V_{np}+V_{\left( \alpha' \right) }=  \notag \\
&=&e^{K}\left[ K^{jk}\left( a_{j}A_{j}a_{k}\bar{A}_{k}e^{-\left(
a_{j}T_{j}+a_{k}\bar{T}_{k}\right) }-\left( a_{j}A_{j}e^{-a_{j}T_{j}}\bar{W}%
K_{k}+a_{k}\bar{A}_{k}e^{-a_{k}\bar{T}_{k}}WK_{j}\right) \right)
\right.
\notag \\
&&\left. +3\xi \frac{\left( \xi ^{2}+7\xi
\mathcal{V}+\mathcal{V}^{2}\right)
}{\left( \mathcal{V}-\xi \right) \left( 2\mathcal{V}+\xi \right) ^{2}}%
\left\vert W\right\vert ^{2}\right] .  \label{scalar}
\end{eqnarray}

\bigskip

\section{General Analysis of the String Loop Corrections}
\label{sec3}

\subsection{String Loop Corrections}

Our discussion of the form of the scalar potential in IIB flux
compactifications has still to include the string loop corrections
$\delta K_{(g_{s})}$. These have been computed in full detail only
for unfluxed toroidal orientifolds in \cite{bhk}. Subsequently the
same collaboration in \cite{bhp} made an educated guess for the
behaviour of these loop corrections for general smooth Calabi-Yau
three-folds by trying to understand how the toroidal calculation
would generalize to the Calabi-Yau case. To be self-contained, we
therefore briefly review the main aspects of the toroidal
orientifold calculation of \cite{bhk}.

%\subsubsection{\protect\bigskip
\medskip

%{\noindent\bf
\subsubsection{Exact calculation: \textit{N=2} $K3\times
T^{2}$ and \textit{N=1 }$T^{6}/(%
%TCIMACRO{\U{2124} }%
%BeginExpansion
\mathbb{Z}
%EndExpansion
_{2}\times
%TCIMACRO{\U{2124} }%
%BeginExpansion
\mathbb{Z}
%EndExpansion
_{2})$} \label{sec3.1.1}

\medskip

The string loop corrections to \textit{N=1 }supersymmetric\textit{\ }$T^{6}/(%
%TCIMACRO{\U{2124} }%
%BeginExpansion
\mathbb{Z}
%EndExpansion
_{2}\times
%TCIMACRO{\U{2124} }%
%BeginExpansion
\mathbb{Z}
%EndExpansion
_{2})$ orientifold compactifications with D3 and D7 branes follow
by generalising the result for \textit{N=2} supersymmetric
$K3\times T^{2}$ orientifolds. Therefore we start by outlining the
result in the second case.

The one-loop corrections to the K\"{a}hler potential from Klein
bottle, annulus and M\"{o}bius strip diagrams are derived by
integrating the one-loop correction to the tree level K\"{a}hler
metric. These corrections are given by 2-point functions and to
derive the corrections $\delta K_{(g_{s})}$ it is sufficient to
compute just one of these correlators and integrate, since all
corrections to the K\"{a}hler metric come from the same $\delta
K_{(g_{s})}$. From \cite{bhk} the one-loop correction to the
2-point function of the complex structure modulus $U$ of $T^{2}$
is given by, dropping numerical factors,
\begin{equation}
\label{correl} \left\langle V_{U}V_{\bar{U}}\right\rangle \sim
-\left( p_{1}\cdot p_{2}\right)
g_{s}^{2}\alpha'^{-4}V_{4}\frac{vol(T^{2})_{s}}{\left( U+
\bar{U}\right) ^{2}}\mathcal{E}_{2}(A_{i},U),
\end{equation}%
where $V_{4}$ is the regulated volume of the \textit{4D} spacetime, $%
vol(T^{2})_{s}$ denotes the volume of $T^{2}$\ in string frame and
$A_{i}$ are open string moduli. The coefficient
$\mathcal{E}_{2}(A_{i},U)$ is a linear combination of
non-holomorphic Eisenstein series $E_{2}(A,U)$ given by
\begin{equation}
E_{2}(A,U)=\sum_{(n,m)\neq
(0,0)}\frac{\hbox{Re}(U)^{2}}{\left\vert
n+mU\right\vert ^{4}}\exp \left[ 2\pi i\frac{A(n+m\bar{U})+\bar{A}(n+mU)}{U+%
\bar{U}}\right] .  \label{eisenstein}
\end{equation}%
The result (\ref{correl}) is converted to Einstein frame through a
Weyl
rescaling%
\begin{equation}
\left\langle V_{U}V_{\bar{U}}\right\rangle _{E}=\left\langle V_{U}V_{\bar{U}%
}\right\rangle _{s}\frac{e^{2\varphi }}{vol(K3\times T^{2})_{s}},
\end{equation}%
giving
\begin{equation}
\left\langle V_{U}V_{\bar{U}}\right\rangle \sim -\left( p_{1}\cdot
p_{2}\right) g_{s}^{2}\alpha'^{-4}V_{4}\frac{e^{2\varphi }}{\left( U+%
\bar{U}\right) ^{2}}\frac{\mathcal{E}_{2}(A_{i},U)}{vol(K3)_{s}}.
\end{equation}%
Writing the volume of the $K3$ hypersurface in Einstein frame
%\begin{equation}
$vol(K3)_{s}=e^{\varphi }vol(K3)_{E},$
%\end{equation}%
%which, in turn,
produces the final result%
\begin{equation}
\left\langle V_{U}V_{\bar{U}}\right\rangle \sim -\left( p_{1}\cdot
p_{2}\right) g_{s}^{2}\alpha'^{-4}V_{4}\frac{e^{\varphi }}{\left( U+%
\bar{U}\right) ^{2}}\frac{\mathcal{E}_{2}(A_{i},U)}{vol(K3)_{E}}.
\label{U}
\end{equation}%
Now noticing that%
\begin{equation}
\partial _{U}\partial _{\bar{U}}E_{2}(A,U)\sim -\frac{E_{2}(A,U)}{\left( U+%
\bar{U}\right) ^{2}},
\end{equation}%
 we can read off from (\ref{U}) the 1-loop correction to the kinetic term for
the field $U$  and using
 $vol(K3)_{E}=\tau $, the 1-loop correction to the K\"{a}hler potential becomes
\begin{equation}
\delta K_{(g_{s})}= c \frac{\mathcal{E}_{2}(A_{i},U)}{
\hbox{Re}\left( S\right) \tau},  \label{UU}
\end{equation}%
where a full analysis determines the constant of proportionality
$c$ to be $c=-1/(128\pi ^{4})$\footnote{The constant $c$ given
here differs from the one calculated in \cite{bhk} only by a
factor of $(-\pi^{2})$ due to different conventions. In fact, in
\cite{bhk} the correction (\ref{UU}) takes the form $\delta
K_{(g_{s})}= -\frac{c}{8}
\frac{\mathcal{E}_{2}(A_{i},U)}{\hbox{Im}( S)\hbox{Im}(T)}$ with
$\hbox{Im}(S)\equiv \frac{e^{-\varphi}}{\sqrt{8}\pi}$ and
$\hbox{Im}(T)\equiv \frac{\tau}{\sqrt{8}\pi}$.}. This procedure
can be generalized to evaluate the loop corrections in the
\textit{N=1} supersymmetric\textit{\ }$T^{6}/(%
%TCIMACRO{\U{2124} }%
%BeginExpansion
\mathbb{Z}
%EndExpansion
_{2}\times
%TCIMACRO{\U{2124} }%
%BeginExpansion
\mathbb{Z}
%EndExpansion
_{2})$ case, obtaining
\begin{equation}
\delta K_{(g_{s})}=\delta K_{(g_{s})}^{KK}+\delta K_{(g_{s})}^{W},
\end{equation}%
where $\delta K_{(g_{s})}^{KK}$ comes from the exchange between D7
and D3-branes of closed strings which carry Kaluza-Klein momentum,
and gives (for vanishing open string scalars)
\begin{equation}
\delta K_{(g_{s})}^{KK}= -\frac{1}{128\pi^{4}}\sum\limits_{i=1}^{3}%
\frac{\mathcal{E}_{i}^{KK}(U,\bar{U})}{\hbox{Re}\left( S\right)
\tau _{i}}. \label{KK}
\end{equation}
The other correction $\delta K_{(g_{s})}^{W}$ can again be
interpreted in the closed string channel as coming from the
exchange of winding strings between intersecting stacks of
D7-branes. These contributions are present in the \textit{N=1}
case but not in the \textit{N=2} case. They
take the form%
\begin{equation}
\delta K_{(g_{s})}^{W}=-\frac{1}{128\pi^{4}}\sum\limits_{i\neq
j\neq k=1}^{3}\frac{\mathcal{E}_{i}^{W}(U,\bar{U})}{\tau _{j}\tau
_{k}}. \label{W}
\end{equation}%

\medskip
%\subsubsection
%{\noindent\bf
\subsubsection{Generalisation to Calabi-Yau three-folds}
\medskip

The previous calculation teaches us that, regardless of the
particular background under consideration, a Weyl rescaling will
always be necessary to convert to four-dimensional Einstein frame.
This implies the 2-point function should always be suppressed by
the overall volume:
\begin{equation}
\left\langle V_{U}V_{\bar{U}}\right\rangle _{s}\sim g(U,T,S)%
\Longleftrightarrow \left\langle V_{U}V_{\bar{%
U}}\right\rangle _{E}\sim g(U,T,S)\frac{e^{\varphi
/2}}{\mathcal{V}_{E}}.
\end{equation}
This allowed \cite{bhp} to conjecture the parametric form of the
loop corrections even for Calabi-Yau cases. $g(U, T, S)$
originates from KK modes as $m_{KK}^{-2}$ and so should scale as a
2-cycle volume $t$. Conversion to Einstein frame then leads to
\begin{equation}
\delta K_{(g_{s})}^{KK}\sim
\sum\limits_{i=1}^{h_{1,1}}g(U)\frac{\left(
a_{l}t^{l}\right) e^{\varphi }}{\mathcal{V}}=\sum\limits_{i=1}^{h_{1,1}}%
\frac{\mathcal{C}_{i}^{KK}(U,\bar{U})\left( a_{il}t^{l}\right) }{\hbox{Re}%
\left( S\right) \mathcal{V}},  \label{UUU}
\end{equation}%
where $a_{l}t^{l}$ is a linear combination of the basis 2-cycle volumes $%
t_{l}$. A similar line of argument for the winding corrections
(where the function $g(U, T, S)$ goes as $m_{W}^{-2} \sim t^{-1}$)
gives
\begin{equation}
\delta K_{(g_{s})}^{W}\sim \sum\limits_{i=1}^{h_{1,1}}\frac{\mathcal{C}%
_{i}^{W}(U,\bar{U})}{\left( a_{il}t^{l}\right) \mathcal{V}}.
\label{UUUU}
\end{equation}%

Notice that $\mathcal{C}^{KK}_i $ and $\mathcal{C}^W_i$ are
unknown functions of the complex structure moduli and therefore
this mechanism is only useful to fix the leading order dependence
on K\"{a}hler moduli. This is similar to the K\"{a}hler potential
for matter fields whose dependence on K\"{a}hler moduli can be
extracted by scaling arguments \cite{ccq}, while the complex
structure dependence is unknown. Fortunately it is the K\"{a}hler
moduli dependence that is more relevant in both cases due to the
fact that complex structure moduli are naturally fixed by fluxes
at tree-level. On the other hand, the K\"{a}hler moduli need
quantum corrections to be stabilised and are usually more relevant
for supersymmetry breaking.

We now turn to trying to understand the loop corrections from a
low-energy point of view.

\subsection{Low Energy Approach}

The low energy physics is described by a four dimensional
supergravity action. We ask here whether it is possible to
understand the form of the loop corrections in terms of the
properties of the low energy theory, without relying on a full
string theory computation.

We first ask what one could reasonably hope to understand.
The form of equations (\ref{KK}) and (\ref{W}) show a very complicated dependence on the
complex structure moduli, and a very simple dependence on the dilaton and K\"ahler moduli.
The dependence on the complex structure moduli is associated with an Eisenstein series originating
from the structure of the torus, and so we cannot expect to reproduce this without a full string computation.
On the other hand the dilaton and K\"ahler moduli appear with a very simple scaling behaviour. This we
may hope to be able to understand using low-energy arguments, and to be able to conjecture the generalisation
to the Calabi-Yau case.

There is one paper in the literature that has already tried to do
that. In an interesting article \cite{hg}, von Gersdorff and
Hebecker considered models with one K\"{a}hler modulus $\tau $,
such that $\mathcal{V}=\tau ^{3/2}=R^{6}$ $\Longleftrightarrow
\tau =R^{4}$, and argued for the form of $\delta K_{(g_{s})}^{KK}$
using the Peccei-Quinn symmetry, scaling arguments and the
assumption that the loop corrections arise simply from the
propagation of $10D$ free fields in the compact space and
therefore do not depend on $M_{s}$. This led to the proposal
\begin{equation}
 \delta K_{(g_{s})}^{KK}\simeq \tau
^{-2}.  \label{von Gersdorff}
\end{equation}
However, at the level of the K\"{a}hler potential (but not the
scalar potential) this result disagrees with the outcome of the
exact toroidal calculation (\ref{KK}). It seems on the contrary to
reproduce the corrections due to the exchange of winding strings
(\ref{W}), but as $m_W > M_s > m_{KK}$ we do not expect to see
such corrections at low energy. In reality, $\ \delta
K_{(g_{s})}^{KK}$ should contain all contributions to the 1-loop
corrections to the kinetic term of $\tau$. From the reduction of
the DBI action we know that $\tau $ couples to the field theory on
the stack of $D7$-branes wrapping the 4-cycle whose volume is
given by $\tau $. It therefore does not seem that the string loop
corrections will come from the propagation of free fields as
$\tau$ will interact with the corresponding gauge theory on the
brane. In fact the reduced DBI action contains a term which looks
like
\begin{equation}
\delta S_{DBI}\supset \int d^{4}x\sqrt{-g^{(4)}}\tau F^{\mu \nu
}F_{\mu \nu },
\end{equation}%
and when $\tau $\ gets a non-vanishing VEV, expanding around this
VEV in the following way
\begin{equation}
\tau =\left\langle \tau \right\rangle +\tau ' ,
\end{equation}
we obtain
\begin{equation}
\delta S_{DBI}\supset \int d^{4}x\sqrt{-g^{(4)}}\left(
\left\langle \tau \right\rangle F^{\mu \nu }F_{\mu \nu }+\tau '
F^{\mu \nu }F_{\mu \nu }\right) .  \label{DBI}
\end{equation}
%\begin{figure}[ht]
%\linespread{0.2}
%\begin{center}
\FIGURE{ \epsfxsize=0.40\hsize \epsfbox{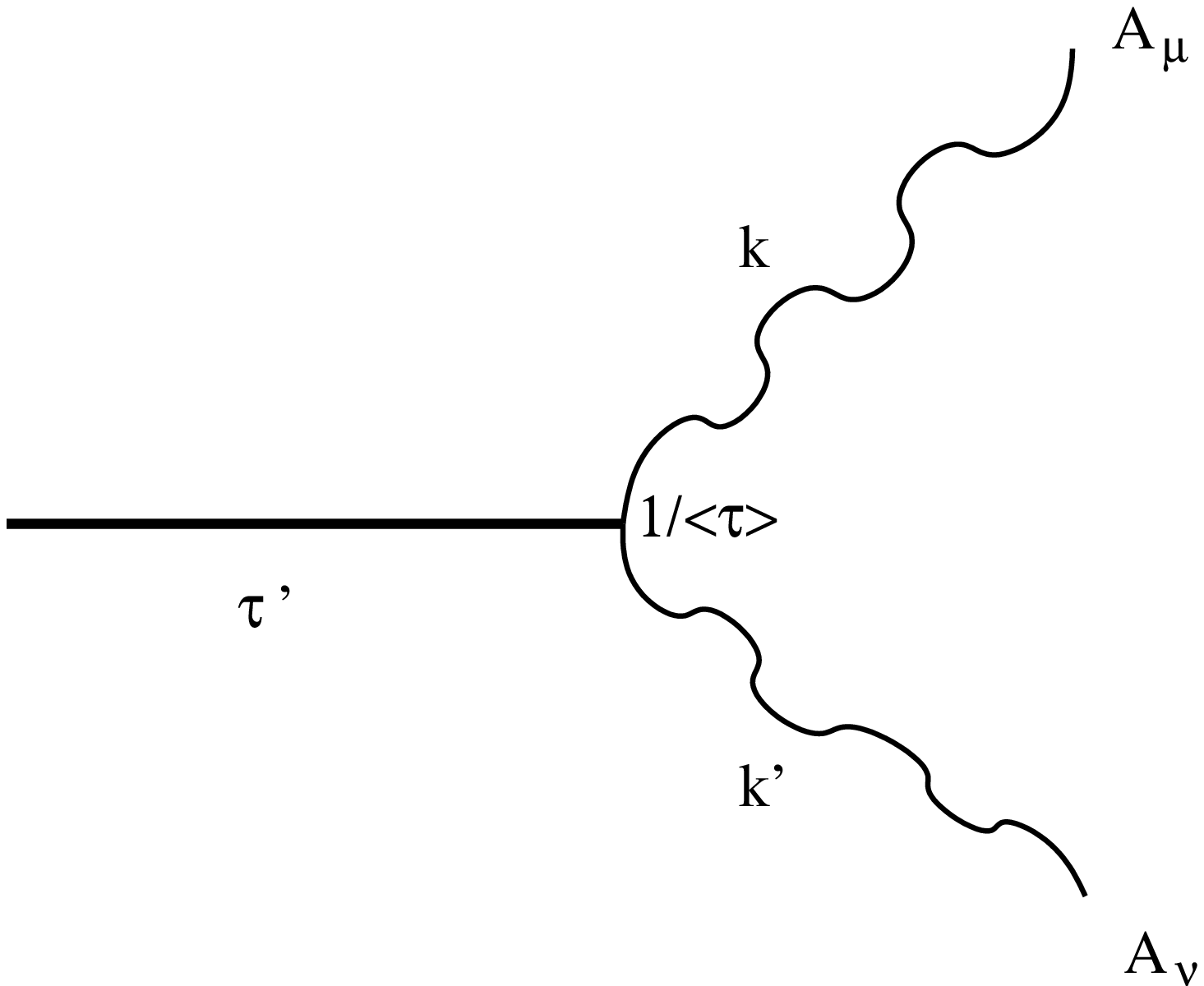}
\caption{Coupling of the K\"ahler modulus with the gauge fields on
the brane.} }
%\end{center}
%\end{figure}
From the first term in (\ref{DBI}) we can readily read off the
coupling constant of the gauge group on the brane
\begin{equation}
g^{2}=\frac{1}{M_{s}^{4}\tau },  \label{coupling}
\end{equation}%
where we have added $M_{s}^{4}$\ to render it correctly
dimensionless. On the other hand, the second term in (\ref{DBI})
will give rise to an interaction vertex of the type shown in
Figure 1 that will affect the 1-loop renormalisation of the $\tau$
kinetic term.

In any ordinary quantum field theory, generic scalar fields
$\varphi $ get 1-loop quantum corrections to their
kinetic terms (wavefunction renormalisation) of the form%
\begin{equation}
\int d^{4}x\sqrt{-g^{(4)}}\frac{1}{2}\left( 1+A\right) \partial
_{\mu }\varphi
\partial ^{\mu }\varphi,  \label{hj}
\end{equation}%
where $A$ is given by $A\simeq \frac{g^{2}}{16\pi ^{2}}$, with $g$
the coupling constant of the gauge interaction this scalar couples
to.

$\tau$ is a modulus and not a gauge-charged field. Nonetheless, we
still expect loop corrections to generate corrections to the
moduli kinetic terms. We expect to be able to write the kinetic
terms as
\begin{equation}
K_{i \bar{j}} = K_{i \bar{j}, tree} + \delta K_{i \bar{j},
1-loop}.
\end{equation}
We also expect the loop correction to the kinetic term to always
involve a suppression by the coupling that controls the loop
expansion. This is the analogue to the correction in (\ref{hj})
depending on the gauge coupling constant, which controls the loop
expansion of ordinary field theory. For a brane wrapping a cycle
$\tau$, the value of $\tau$ is the gauge coupling for branes
wrapping the cycle, and we expect loop corrections involving those
branes to involve a suppression, relative to tree-level terms, by
a factor of $\tau$ (see \cite{joe} for related arguments).

This is not a rigorous derivation, but we consider this a reasonable assumption. We will find that it gives the correct
scaling of the loop correction for the toroidal case where the correction has been computed explicitly, and that, while
it has a different origin,
it agrees with the BHP conjecture for the parametric form of loop corrections in the Calabi-Yau case.
The loop corrections to the K\"ahler potential $K$ should then be such as to
generate corrections to the kinetic terms for $\tau$ that are
suppressed by a factor of $g^2$ for the gauge theory on branes wrapping the cycle $\tau$.
The
K\"{a}hler potential upon double differentiation yields
the kinetic terms in the \textit{4D} Einstein frame Lagrangian%
\begin{equation}
S_{\text{\textit{Einstein}}}\supset \int d^{4}x\sqrt{-g^{(4)}}\left( \frac{%
\partial ^{2}\left( K_{tree}\right) }{\partial \tau ^{2}}+\frac{\partial
^{2}\left( \delta K_{(g_{s})}^{KK}\right) }{\partial \tau
^{2}}\right) \left( \partial \tau \right) ^{2},
\end{equation}%
and the general canonical redefinition of the scalar fields%
\begin{equation}
\tau \longrightarrow \varphi =\varphi (\tau ),
\end{equation}%
will produce a result similar to (\ref{hj}), which implies%
\begin{equation}
\frac{\partial ^{2}\left( K_{tree}\right) }{\partial \tau ^{2}}%
\longrightarrow \frac{1}{2},\text{ \ \ \ \ }\frac{\partial
^{2}\left( \delta
K_{(g_{s})}^{KK}\right) }{\partial \tau ^{2}}\longrightarrow \frac{1}{2}%
A\sim \frac{1}{2}\frac{g^{2}}{16\pi ^{2}},
\end{equation}%
and thus%
\begin{equation}
\frac{\partial ^{2}\left( \delta K_{(g_{s})}^{KK}\right) }{\partial \tau ^{2}%
}\sim \frac{g^{2}}{16\pi ^{2}}\frac{\partial ^{2}\left(
K_{tree}\right) }{\partial \tau ^{2}}.  \label{dds}
\end{equation}%
Using  equation (\ref{coupling}) we then guess for the
scaling behavior of the string loop corrections to the K\"{a}hler potential%
\begin{equation}
\frac{\partial ^{2}\left( \delta K_{(g_{s})}^{KK}\right) }{\partial \tau ^{2}%
}\sim \frac{f(\hbox{Re}(S))}{16\pi ^{2}}\frac{1}{\tau
}\frac{\partial ^{2}\left( K_{tree}\right) }{\partial \tau ^{2}},
\label{fundamentale}
\end{equation}%
where we have introduced an unknown function of the dilaton
$f(\hbox{Re}(S))$ representing an integration constant\footnote{In
general there should be also an unknown function of the complex
structure and open string moduli but we dropped it since, as we
stated at the beginning of this section, its full determination
would require an exact string calculation.}. However we may be
able to use similar reasoning to determine $f(\hbox{Re}(S))$. The
same correction $\delta K_{(g_{s})}^{KK}$, upon double
differentiation with respect to the dilaton, has to give rise to
the 1-loop quantum correction to the corresponding dilaton kinetic
term. We also recall that $S$ couples to all field theories on
$D3$-branes as the relative gauge kinetic function is the dilaton
itself. Using the same argument as above we end up with the
further guess for $\delta K_{(g_{s})}^{KK}$:
\begin{equation}
\frac{\partial ^{2}\left( \delta K_{(g_{s})}^{KK}\right) }{\partial \hbox{Re}%
(S)^{2}}\sim \frac{h(\tau )}{16\pi
^{2}}\frac{1}{\hbox{Re}(S)}\frac{\partial ^{2}\left(
K_{tree}\right) }{\partial \hbox{Re}(S)^{2}}\simeq \frac{h(\tau
)}{16\pi ^{2}}\frac{1}{\hbox{Re}(S)^{3}},  \label{jkj}
\end{equation}%
where $h(\tau )$ is again an unknown function which parameterises
the
dependence on the K\"{a}hler modulus. Integrating (\ref{jkj}) twice, we obtain%
\begin{equation}
\delta K_{(g_{s})}^{KK}\sim \frac{h(\tau )}{16\pi ^{2}}\frac{1}{\hbox{Re}(S)}%
,  \label{dilatonic}
\end{equation}%
where $h(\tau )$ can be worked out from (\ref{fundamentale})%
\begin{equation}
\frac{\partial ^{2}\left( h(\tau )\right) }{\partial \tau ^{2}}\sim \frac{1}{%
\tau }\frac{\partial ^{2}\left( K_{tree}\right) }{\partial \tau
^{2}}. \label{fundamental}
\end{equation}%

We now apply the above methods to several
Calabi-Yau cases, comparing to either the exact results or the
conjecture of equation (\ref{UUU})

\subsubsection{Case 1: \textit{N=1 }$T^{6}/(%
%TCIMACRO{\U{2124} }%
%BeginExpansion
\mathbb{Z}
%EndExpansion
_{2}\times
%TCIMACRO{\U{2124} }%
%BeginExpansion
\mathbb{Z}
%EndExpansion
_{2})$} \label{sec3.2.1}

We first consider the case of toroidal compactifications, for
which the loop corrections have been explicitly computed
\cite{bhk}. In that case the volume can be expressed as (ignoring
the 48 twisted K\"{a}hler moduli obtained by blowing up
orbifold singularities)%
\begin{equation}
\mathcal{V}=\sqrt{\tau _{1}\tau _{2}\tau _{3}},  \label{torus}
\end{equation}%
and so (\ref{fundamental})\ takes the form%
\begin{equation}
\frac{\partial ^{2}\left( \delta K_{(g_{s})}^{KK}\right)
}{\partial \tau _{i}^{2}}\sim \frac{f(\hbox{Re}(S))}{16\pi
^{2}}\frac{1}{\tau _{i}^{3}}\text{ \ \ \ \ \ }\forall i=1,2,3.
\end{equation}%
Upon integration we get%
\begin{equation}
\delta K_{(g_{s})}^{KK}\sim \frac{1}{16\pi
^{2}}\frac{f(\hbox{Re}(S))}{\tau _{i}}\text{ \ \ \ \ \ }\forall
i=1,2,3.
\end{equation}%
Now combining this result with the analysis for the dilatonic
dependence of
the string loop corrections, we obtain%
\begin{equation}
\delta K_{(g_{s})}^{KK}\sim \frac{1}{16\pi
^{2}}\sum_{i=1}^{3}\frac{1}{\hbox{Re}(S)\tau _{i}},
\end{equation}%
which reproduces the scaling behaviour of the result (\ref{KK})\
found from string scattering amplitudes.

\subsubsection{Case 2: $%
%TCIMACRO{\U{2102} }%
%BeginExpansion
\mathbb{C}
%EndExpansion
P_{[1,1,1,6,9]}^{4}$} \label{sec3.2.2}

We next consider loop corrections to the K\"{a}hler potential for
the Calabi-Yau orientifold $\mathbb{C} P_{[1,1,1,6,9]}^{4}$. We
will compare the form of (\ref{UUU}) (see also (\ref{AA})) to that
arising from our method (\ref{fundamental}) to work out the
behaviour of $\delta K_{(g_{s})}^{KK}$,
finding again a matching.\footnote{We note that the topology of $%
\mathbb{C} P_{[1,1,1,6,9]}^{4}$ does not allow to have $\delta
K_{(g_{s})}^{W}\neq 0$ \cite{curio}.}
In the large volume limit we can write the volume %\ (\ref{volume})\
as follows
\begin{equation}
\mathcal{V}=\frac{1}{9\sqrt{2}}\left( \tau _{5}^{3/2}-\tau
_{4}^{3/2}\right) \simeq \tau _{5}^{3/2},
\end{equation}%
and (\ref{UUU}) becomes
\begin{equation}
\delta K_{(g_{s})}^{KK}\sim \frac{\mathcal{C}_{4}^{KK}\sqrt{\tau _{4}}}{%
\hbox{Re}\left( S\right)
\mathcal{V}}+\frac{\mathcal{C}_{5}^{KK}\sqrt{\tau
_{5}}}{\hbox{Re}\left( S\right) \mathcal{V}}\simeq \frac{\mathcal{C}_{4}^{KK}%
\sqrt{\tau _{4}}}{\hbox{Re}\left( S\right) \mathcal{V}}+\frac{\mathcal{C}%
_{5}^{KK}}{\hbox{Re}\left( S\right) \tau _{5}}.  \label{hjkl}
\end{equation}%
From the tree-level K\"{a}hler matrix we read%
\begin{equation}
\frac{\partial ^{2}\left( K_{tree}\right) }{\partial \tau
_{4}^{2}}\simeq \frac{1}{\sqrt{\tau _{4}}\mathcal{V}},\text{ \ \ \
\ }\frac{\partial ^{2}\left( K_{tree}\right) }{\partial \tau
_{5}^{2}}\simeq \frac{1}{\tau _{5}^{2}}.
\end{equation}%
Requiring loop corrections to be suppressed by a factor of
$g^2_{c}$ for the field-theory on the brane gives
\begin{equation}
\left\{
\begin{array}{c}
\frac{\partial ^{2}\left( \delta K_{(g_{s})}^{KK}\right)
}{\partial \tau _{4}^{2}}\sim \frac{1}{16\pi
^{2}}\frac{1}{\hbox{Re}(S)}\frac{1}{\tau
_{4}^{3/2}\mathcal{V}} \\
\frac{\partial ^{2}\left( \delta K_{(g_{s})}^{KK}\right)
}{\partial \tau _{5}^{2}}\sim
\frac{1}{16\pi ^{2}}\frac{1}{\hbox{Re}(S)}\frac{1}{\tau _{5}^{3}}%
\end{array}%
\right.
\end{equation}%
which, upon double integration, matches exactly the scaling
behaviour of the result (\ref{hjkl}).

\subsubsection{Case 3: $%
%TCIMACRO{\U{2102} }%
%BeginExpansion
\mathbb{C}
%EndExpansion
P_{[1,1,2,2,6]}^{4}$} \label{sec3.2.3}

\bigskip
As another example we study the expected form of loop corrections
for the case of the Calabi-Yau manifold $\mathbb{C}
P_{[1,1,2,2,6]}^{4}$, defined by the degree 12 hypersurface
embedding. This Calabi-Yau is a K3 fibration and has $(h^{1,1},
h^{2,1})=(2,128)$ with $\chi =-252$. Including only the complex
structure deformations that survive the mirror map, the
defining equation is%
\begin{equation}
z_{1}^{12}+z_{2}^{12}+z_{3}^{6}+z_{4}^{6}+z_{5}^{2}-12\psi
z_{1}z_{2}z_{3}z_{4}z_{5}-2\phi z_{1}^{6}z_{2}^{6}=0.
\end{equation}%
In terms of 2-cycle volumes the overall volume takes the form
\begin{equation}
\mathcal{V}=t_{1}t_{2}^{2}+\frac{2}{3}t_{2}^{3},
\end{equation}%
giving relations between the 2- and 4-cycle volumes, \bea
\label{tay} \tau _{1}=t_{2}^{2}, & \qquad &
\tau_{2}=2t_{2}\left(t_{1}+t_{2}\right),
\nonumber \\
t_{2}=\sqrt{\tau _{1}}, & \qquad & t_{1}=\frac{\tau _{2}-2\tau
_{1}}{2\sqrt{\tau _{1}}}, \eea allowing us to write
\begin{equation}
\mathcal{V}=\frac{1}{2}\sqrt{\tau _{1}}\left( \tau
_{2}-\frac{2}{3}\tau _{1}\right) .  \label{vol11226}
\end{equation}%

Let us now investigate what the arguments above suggest for the form of the
string loop corrections for the $%
%TCIMACRO{\U{2102} }%
%BeginExpansion
\mathbb{C}
%EndExpansion
P_{[1,1,2,2,6]}^{4}$ model should look like. Applying (\ref{UUU})
and (\ref{UUUU}) for the one-loop correction to
$K$, we find%
\begin{equation}
\delta K_{(g_{s})}^{KK}\sim
\frac{\mathcal{C}_{1}^{KK}}{\hbox{Re}\left(
S\right) \mathcal{V}}\frac{\tau _{2}-2\tau _{1}}{2\sqrt{\tau _{1}}}+\frac{%
\mathcal{C}_{2}^{KK}\sqrt{\tau _{1}}}{\hbox{Re}\left( S\right)
\mathcal{V}}, \label{ggg}
\end{equation}%
along with%
\begin{equation}
\delta K_{(g_{s})}^{W}\sim \frac{\mathcal{C}_{1}^{W}}{\mathcal{V}}\frac{2%
\sqrt{\tau _{1}}}{\tau _{2}-2\tau _{1}}+\frac{\mathcal{C}_{2}^{W}}{\mathcal{V%
}\sqrt{\tau _{1}}}.  \label{gggg}
\end{equation}%
The arguments summarized in the relation (\ref{fundamental})
reproduce exactly the behaviour of these
corrections. The tree-level K\"{a}hler metric reads%
\begin{equation}
\frac{\partial ^{2}\left( K_{tree}\right) }{\partial \tau _{1}^{2}}=\frac{1}{%
\tau _{1}^{2}}+\frac{2}{9}\frac{\tau _{1}}{\mathcal{V}^{2}},\text{ \ \ \ \ }%
\frac{\partial ^{2}\left( K_{tree}\right) }{\partial \tau _{2}^{2}}=\frac{1}{%
2}\frac{\tau _{1}}{\mathcal{V}^{2}}. \label{matrixElem}
\end{equation}%
Given that we are interested simply in the scaling behaviour of
these corrections, we notice that either in the case $\tau_{1}
\lesssim \tau_{2}$ such that
\begin{equation}
\mathcal{V}=\frac{1}{2}\sqrt{\tau _{1}}\left( \tau
_{2}-\frac{2}{3}\tau _{1}\right) \simeq \tau _{1}^{3/2}\simeq \tau
_{2}^{3/2},  \label{ooo}
\end{equation}%
or in the large volume limit $\tau_{1} \ll \tau_{2}$ where%
\begin{equation}
\mathcal{V} \simeq \sqrt{\tau _{1}}\tau _{2},
\end{equation}
the matrix elements (\ref{matrixElem}) take the form
\begin{equation}
\frac{\partial ^{2}\left( K_{tree}\right) }{\partial \tau
_{1}^{2}}\sim
\frac{1}{\tau _{1}^{2}},\ \ \ \ \frac{\partial ^{2}\left( K_{tree}\right) }{%
\partial \tau _{2}^{2}}\sim \frac{1}{\tau _{2}^{2}}.
\end{equation}%
We can now see that our method (\ref{fundamental})\ yields

\begin{equation}
\left\{
\begin{array}{c}
\frac{\partial ^{2}\left( \delta K_{(g_{s})}^{KK}\right)
}{\partial \tau
_{1}^{2}}\sim \frac{1}{16\pi ^{2}}\frac{1}{\hbox{Re}(S)\tau _{1}}\frac{%
\partial ^{2}\left( K_{tree}\right) }{\partial \tau _{1}^{2}}%
\Longleftrightarrow \delta K_{(g_{s},\tau _{1})}^{KK}\sim \frac{1}{\hbox{Re}%
(S)\tau _{1}} \\
\frac{\partial ^{2}\left( \delta K_{(g_{s})}^{KK}\right)
}{\partial \tau _{2}^{2}}\sim \frac{1}{16\pi
^{2}}\frac{1}{\hbox{Re}(S)\tau _{2}}\frac{\partial ^{2}\left(
K_{tree}\right) }{\partial \tau _{2}^{2}}\Longleftrightarrow
\delta
K_{(g_{s},\tau _{2})}^{KK}\sim \frac{1}{\hbox{Re}(S)\tau _{2}}%
\end{array}%
\right.
\end{equation}%
which, both in the case $\tau_{1} \lesssim \tau_{2}$ and $\tau_{1}
\ll \tau_{2}$, matches the scaling
behaviour of (\ref{ggg})%
\begin{equation}
\delta K_{(g_{s})}^{KK}\sim
\frac{\mathcal{C}_{1}^{KK}}{\hbox{Re}\left(
S\right) \mathcal{V}}\frac{\tau _{2}-2\tau _{1}}{2\sqrt{\tau _{1}}}+\frac{%
\mathcal{C}_{2}^{KK}\sqrt{\tau _{1}}}{\hbox{Re}\left( S\right) \mathcal{V}}%
\sim \frac{\mathcal{C}_{1}^{KK}}{\hbox{Re}\left( S\right) \tau _{1}}+\frac{%
\mathcal{C}_{2}^{KK}}{\hbox{Re}\left( S\right) \tau _{2}}.
\end{equation}

\section{Extended No Scale Structure}
\label{sec4}

The examples in the previous section give support to the notion
that loop corrections to the K\"{a}hler potential can be
understood by requiring that the loop-corrected kinetic terms for a modulus $\tau$ are
suppressed by a factor of $g^2$ for the gauge group on branes
wrapping the $\tau$ cycle. We repeat again that these arguments only apply to moduli that control loop factors.

While not proven, we now assume the validity of this parametric form of the corrections and
move on to study the effect of such corrections in the
scalar potential. We shall show that the leading contribution to
the scalar potential is null, due to a cancellation in the expression for the scalar potential.
 We shall see that this cancellation holds so long as $\delta
K_{(g_{s})}^{KK}$ is an homogeneous function of degree $n=-2$ in
the 2-cycle volumes. We call this ``extended no scale structure",
as the cancellation in the scalar potential that is characteristic
of no-scale models extends to one further order, so that compared
to a naive expectation the scalar potential is only non-vanishing
at sub-sub-leading order. Let us state clearly the ``extended
no-scale structure" result:

\begin{quotation}
\textit{Let $X$ be a Calabi-Yau three-fold and consider type IIB
$N=1$ $4D$ $SUGRA$ where the K\"{a}hler potential and the
superpotential in the Einstein frame take the form:%
\begin{equation}
\left\{
\begin{array}{l}
K=K_{tree}+\delta K,\\
W=W_{0}.
\end{array}%
\right.
\end{equation}%
If and only if the loop correction $\delta K$ to $K$ is a
homogeneous function
in the 2-cycles volumes of degree $n=-2$, then at leading order}%
\begin{equation}
\delta V_{(g_{s})}=0.
\end{equation}
\end{quotation}
We shall provide now a rigorous proof of the previous claim. We
are interested only in the perturbative part of the scalar
potential. We therefore
 focus on%
\begin{equation}
\delta V_{(g_{s})}=\left( K^{ij}\partial _{i}K\partial _{j}K-3\right) \frac{%
\left\vert W\right\vert ^{2}}{\mathcal{V}^{2}},  \label{pp}
\end{equation}%
where $K=-2\ln \left( \mathcal{V}\right) +\delta K_{(g_{s})}$. We
focus on $\delta K$ coming from $g_s$ (rather than $\alpha'$)
corrections. We require the inverse of the quantum corrected
K\"{a}hler matrix, which can be found using the Neumann series.
Introducing an expansion parameter $\varepsilon $, and writing
$K_{tree}$ as $K_{0}$, we define
\begin{equation}
\mathcal{K}_{0}=\left\{ \frac{\partial ^{2}K_{0}}{\partial \tau
_{i}\partial
\tau _{j}}\right\} _{i,j=1,...,h_{1,1}},\text{ \ \ \ \ \ }\delta \mathcal{K}%
=\left\{ \frac{\partial ^{2}\left( \delta K_{(g_{s})}\right)
}{\partial \tau _{i}\partial \tau _{j}}\right\}
_{i,j=1,...,h_{1,1}}
\end{equation}%
and have
\begin{equation}
K^{ij}=\left( \mathcal{K}_{0}+\varepsilon \delta
\mathcal{K}\right)
^{ij}=\left( \mathcal{K}_{0}\left( \mathbf{1}+\varepsilon \mathcal{K}%
_{0}^{-1}\delta \mathcal{K}\right) \right) ^{ij}=\left( \mathbf{1}%
+\varepsilon \mathcal{K}_{0}^{-1}\delta \mathcal{K}\right)
^{il}K_{0}^{lj}.
\end{equation}%
Now use the Neumann series%
\begin{equation}
\left( \mathbf{1}+\varepsilon \mathcal{K}_{0}^{-1}\delta
\mathcal{K}\right) ^{il}=\delta _{l}^{i}-\varepsilon
K_{0}^{im}\delta K_{ml}+\varepsilon ^{2}K_{0}^{im}\delta
K_{mp}K_{0}^{pq}\delta K_{ql}+\mathcal{O}(\varepsilon ^{3}),
\end{equation}%
to find%
\begin{equation}
K^{ij}=K_{0}^{ij}-\varepsilon K_{0}^{im}\delta
K_{ml}K_{0}^{lj}+\varepsilon
^{2}K_{0}^{im}\delta K_{mp}K_{0}^{pq}\delta K_{ql}K_{0}^{lj}+\mathcal{O}%
(\varepsilon ^{3}).  \label{ppp}
\end{equation}%
Substituting (\ref{ppp}) back in (\ref{pp}), we obtain%
\begin{equation}
\delta V_{(g_{s})}=V_{0}+\varepsilon \delta V_{1}+\varepsilon
^{2}\delta V_{2}+\mathcal{O}(\varepsilon ^{3}), \label{expansion
of V}
\end{equation}%
where $V_{0}=\left( K_{0}^{ij}K_{i}^{0}K_{j}^{0}-3\right)
\frac{\left\vert W\right\vert ^{2}}{\mathcal{V}^{2}}=0$ due to
(\ref{no scale}) is the usual no-scale structure and
\begin{equation}
\left\{
\begin{array}{c}
\delta V_{1}=\left( 2K_{0}^{ij}K_{i}^{0}\delta
K_{j}-K_{0}^{im}\delta
K_{ml}K_{0}^{lj}K_{i}^{0}K_{j}^{0}\right) \frac{\left\vert W\right\vert ^{2}%
}{\mathcal{V}^{2}} \\
\delta V_{2}=\left( K_{0}^{ij}\delta K_{i}\delta
K_{j}-2K_{0}^{im}\delta
K_{ml}K_{0}^{lj}K_{i}^{0}\delta K_{j}\right.  \\
\left. +K_{0}^{im}\delta K_{mp}K_{0}^{pq}\delta
K_{ql}K_{0}^{lj}K_{i}^{0}K_{j}^{0}\right) \frac{\left\vert W\right\vert ^{2}%
}{\mathcal{V}^{2}}.%
\end{array}%
\right.   \label{pppp}
\end{equation}%
We caution the reader that (\ref{expansion of V}) is not a loop
expansion of the scalar potential but rather an expansion of the
scalar potential arising from the
1-loop quantum corrected K\"{a}hler metric. The statement of extended no-scale
structure is that $\delta V_{1}$ will vanish, while
 $\delta V_{2}$ will be non-vanishing.
Recalling (\ref{p1}), $\delta V_{1}$\ simplifies to%
\begin{equation}
\delta V_{1}=-\left( 2\tau _{j}\frac{\partial \left( \delta K\right) }{%
\partial \tau _{j}}+\tau _{m}\tau _{l}\frac{\partial ^{2}\left( \delta
K\right) }{\partial \tau _{m}\partial \tau _{l}}\right)
\frac{\left\vert W\right\vert ^{2}}{\mathcal{V}^{2}}.
\end{equation}%
Let us make a change of
coordinates and work with the 2-cycle volumes instead of the
4-cycles. Using the second of the relations
(\ref{useful}), we deduce%
\begin{equation}
2\tau _{j}\frac{\partial }{\partial \tau _{j}}=t_{l}\frac{\partial }{%
\partial t_{l}},
\end{equation}%
and
\begin{equation}
\tau _{m}\tau _{l}\frac{\partial ^{2}}{\partial \tau _{m}\partial \tau _{l}}=%
\frac{1}{4}t_{i}t_{k}\frac{\partial ^{2}}{\partial t_{i}\partial t_{k}}+%
\frac{1}{4}A_{li}t_{i}t_{k}\frac{\partial \left( A^{lp}\right)
}{\partial t_{k}}\frac{\partial }{\partial t_{p}}.
\end{equation}%
From the definition (\ref{aa}) of $A_{li}$, we notice that
$A_{li}$ is an homogeneous function of degree $n=1$ $\forall l,i$.
Inverting the matrix, we still get homogeneous matrix elements but
now of degree $n=-1$. Finally the Euler theorem for homogeneous
functions, tells us that
\begin{equation}
t_{k}\frac{\partial \left( A^{lp}\right) }{\partial t_{k}}=\left(
-1\right) A^{lp},
\end{equation}%
which gives%
\begin{equation}
\tau _{m}\tau _{l}\frac{\partial ^{2}}{\partial \tau _{m}\partial \tau _{l}}=%
\frac{1}{4}t_{i}t_{k}\frac{\partial ^{2}}{\partial t_{i}\partial t_{k}}-%
\frac{1}{4}t_{p}\frac{\partial }{\partial t_{p}},
\end{equation}%
and, in turn%
\begin{equation}
\delta V_{1}=-\frac{1}{4}\left( 3t_{l}\frac{\partial \left( \delta
K\right)
}{\partial t_{l}}+t_{i}t_{k}\frac{\partial ^{2}\left( \delta K\right) }{%
\partial t_{i}\partial t_{k}}\right) \frac{\left\vert W\right\vert ^{2}}{%
\mathcal{V}^{2}}.
\end{equation}
The form of equation (\ref{UUU}) suggests that for arbitrary
Calabi-Yaus the string loop corrections to $K$ will be homogeneous
functions of the 2-cycle volumes, and in particular that the
leading correction will be of degree $-2$ in 2-cycle volumes.
Therefore if the degree of $\delta K$ is $n
$, the Euler theorem tells us that%
\begin{equation}
\delta V_{1}=-\frac{\left\vert W\right\vert ^{2}}{\mathcal{V}^{2}}\frac{1}{4}%
\left( 3n+n(n-1)\right) \delta K=-\frac{\left\vert W\right\vert ^{2}}{%
\mathcal{V}^{2}}\frac{1}{4}n(n+2)\delta K.  \label{ghj}
\end{equation}%
It follows then, as we claimed above, that only $n=-2$ implies
$\delta V_{1}=0$. In particular, from the conjectures (\ref{UUU})
and (\ref{UUUU}), we see that
\begin{equation}
\left\{
\begin{array}{c}
n=-2\text{ \ for \ }\delta K_{(g_{s})}^{KK}, \\
n=-4\text{ \ for \ }\delta K_{(g_{s})}^{W},
\end{array}%
\right.
\end{equation}%
and so%
\begin{equation}
\left\{
\begin{array}{c}
\delta V_{(g_{s}),1}^{KK}=0, \\
\delta V_{(g_{s}),1}^{W}=-2\delta K_{(g_{s})}^{W}\frac{\left\vert
W\right\vert ^{2}}{\mathcal{V}^{2}}.%
\end{array}%
\right.   \label{final}
\end{equation}

\subsection{General Formula for the Effective Scalar Potential}
\label{sec4.1}

\bigskip Let us now work out the general formula for the effective scalar
potential evaluating also the first non-vanishing contribution of
$\delta
K_{(g_{s})}^{KK}$, that is the $\varepsilon ^{2}$ terms (\ref{pppp})\ in $V$%
\begin{eqnarray}
\delta V_{2} &=&\left( K_{0}^{ij}\delta K_{i}\delta
K_{j}-2K_{0}^{im}\delta
K_{ml}K_{0}^{lj}K_{i}^{0}\delta K_{j}\right.   \notag \\
&&\left. +K_{0}^{im}\delta K_{mp}K_{0}^{pq}\delta
K_{ql}K_{0}^{lj}K_{i}^{0}K_{j}^{0}\right) \frac{\left\vert W\right\vert ^{2}%
}{\mathcal{V}^{2}}.
\end{eqnarray}%
Using (\ref{p1}), $\delta V_{2}$\ simplifies to%
\begin{equation}
\delta V_{2}=\left( K_{0}^{ij}\delta K_{i}\delta K_{j}+2\tau
_{m}\delta K_{ml}K_{0}^{lj}\delta K_{j}+\tau _{m}\tau _{q}\delta
K_{ml}K_{0}^{lp}\delta K_{pq}\right) \frac{\left\vert W\right\vert
^{2}}{\mathcal{V}^{2}}. \label{zz}
\end{equation}%
We now stick to the case where $\delta K_{(g_{s})}^{KK}$ is given
by the conjecture (\ref{UUU}). Considering just the contribution
from one modulus (as the contributions from different terms are
independent), and dropping the
dilatonic dependence, we have%
\begin{equation}
\delta K\rightarrow \delta K_{(g_{s}),\tau _{a}}^{KK}\sim \frac{\mathcal{C}%
_{a}^{KK}t_{a}}{\mathcal{V}}.  \label{uhu}
\end{equation}%
From (\ref{uhu}) we notice that%
\begin{eqnarray}
\delta K_{m} &=&A^{mj}\frac{\partial \left( \delta K\right) }{\partial t^{j}}%
=\mathcal{C}_{a}^{KK}A^{mj}\left( -\frac{t_{a}}{\mathcal{V}^{2}}\frac{%
\partial \left( \mathcal{V}\right) }{\partial t^{j}}+\frac{\delta _{aj}}{%
\mathcal{V}}\right)   \label{zzz} \\
&=&\mathcal{C}_{a}^{KK}\left( -\frac{1}{2}\frac{t_{a}t_{m}}{\mathcal{V}^{2}}+%
\frac{A^{am}}{\mathcal{V}}\right)
=-\mathcal{C}_{a}^{KK}K_{am}^{0},
\end{eqnarray}%
thus%
\begin{equation}
K_{0}^{ij}\delta K_{j}=-\mathcal{C}_{a}^{KK}K_{0}^{ij}K_{aj}^{0}=-\mathcal{C}%
_{a}^{KK}\delta _{ai}.
\end{equation}%
With this consideration (\ref{zz}) becomes%
\begin{equation}
\delta V_{2}=\left( -\mathcal{C}_{a}^{KK}\delta K_{a}-2\mathcal{C}%
_{a}^{KK}\tau _{m}\delta K_{ma}+\tau _{m}\tau _{q}\delta
K_{ml}K_{0}^{lp}\delta K_{pq}\right) \frac{\left\vert W\right\vert ^{2}}{%
\mathcal{V}^{2}}.
\end{equation}%
We need now to evaluate%
\begin{equation}
\tau _{m}\delta K_{ml}=\frac{1}{2}t_{p}\frac{\partial }{\partial t_{p}}%
\left( A^{li}\frac{\partial \left( \delta K\right) }{\partial t_{i}}\right) =%
\frac{1}{2}t_{p}\frac{\partial }{\partial t_{p}}\left( \delta
K_{l}\right) =-2\delta K_{l},
\end{equation}%
that yields%
\begin{eqnarray}
\delta V_{2} &=&\left( -\mathcal{C}_{a}^{KK}\delta K_{a}+4\mathcal{C}%
_{a}^{KK}\delta K_{a}+4\delta K_{l}K_{0}^{lp}\delta K_{p}\right) \frac{%
\left\vert W\right\vert ^{2}}{\mathcal{V}^{2}}= \\
&=&\left( -\mathcal{C}_{a}^{KK}\delta
K_{a}+4\mathcal{C}_{a}^{KK}\delta
K_{a}-4\mathcal{C}_{a}^{KK}\delta K_{a}\right) \frac{\left\vert
W\right\vert
^{2}}{\mathcal{V}^{2}} \\
&=&-\mathcal{C}_{a}^{KK}\delta K_{a}\frac{\left\vert W\right\vert ^{2}}{%
\mathcal{V}^{2}}.
\end{eqnarray}%
With the help of the relation (\ref{zzz}) and replacing the
dilatonic dependence, we can write the previous expression in
terms of the tree-level
K\"{a}hler metric%
\begin{equation}
\delta V_{2}=\frac{\left( \mathcal{C}_{a}^{KK}\right) ^{2}}{\hbox{Re}(S)^{2}}%
K_{aa}^{0}\frac{\left\vert W\right\vert ^{2}}{\mathcal{V}^{2}}.
\label{zzzz}
\end{equation}%
Putting together (\ref{final}) and (\ref{zzzz}), we can now  write
the quantum correction to the scalar potential at leading
order at 1
loop for general Calabi-Yaus, in terms of the cycles $i$ wrapped by the branes and the
quantum corrections to the K\"ahler potential,
\begin{equation}
\frame{$\delta V_{\left( g_{s}\right)
}^{1loop}=\sum\limits_{i}\left( \frac{\left( \mathcal{C}%
_{i}^{KK}\right) ^{2}}{\hbox{Re}(S)^{2}}K_{ii}^{0}-2\delta
K_{(g_{s}),\tau _{i}}^{W}\right)
\frac{W_{0}^{2}}{\mathcal{V}^{2}}$}.  \label{1 loop}
\end{equation}
We emphasise that this formula assumes the validity of the BHP conjecture, and only focuses on corrections of this
nature.

Finally we point out that, due to the extended no-scale structure,
in the presence of non-perturbative contributions to the
superpotential, it is also important to check that the leading
quantum corrections to the general scalar potential (\ref{scalar})
are indeed given by (\ref{1 loop}) and the contribution to the
non-perturbative part of the scalar potential generated by string
loop corrections
\begin{equation}
\delta V_{np}=\left(2 K_{0}^{ij}W _{i}\delta K_{(g_{s}),j} W +
\delta
K_{(g_{s})}^{ij}W _{i}W _{j} \right) \frac{%
\left\vert W\right\vert ^{2}}{\mathcal{V}^{2}},  \label{dVnp}
\end{equation}
is irrelevant. A quick calculation shows that this is indeed the
case.\footnote{We shall not discuss the effects of higher loop
contributions to the scalar potential. We expect that these will
be suppressed compared to the one-loop contribution by additional
loop factors of $(16 \pi^2)$, and so will not compete with the
terms considered in (\ref{1 loop}).}

\subsection{Field Theory Interpretation}

We now interpret the above arguments and in particular the existence
of the extended no-scale structure in light of the
Coleman-Weinberg potential \cite{cw}.\footnote{For a previous attempt at matching
string effective actions onto the Coleman-Weinberg potential, see \cite{hepth0003185}.}
 We will see that this gives
a quantitative explanation for the cancellation that is present.
The Coleman-Weinberg potential is given in supergravity by (e.g.
see \cite{fkz})
\begin{equation}
\delta V_{1loop}=\frac{1}{64\pi ^{2}}\left[ \Lambda ^{4} STr\left(
M^{0}\right) \ln \left( \frac{\Lambda ^{2}}{\mu ^{2}}\right)
+2\Lambda ^{2} STr\left( M^{2}\right) + STr \left( M^{4}\ln \left(
\frac{M^{2}}{\Lambda ^{2}}\right) \right) \right] ,
\label{Coleman}
\end{equation}%
where $\mu $ is a scale parameter, $\Lambda$ the cut-off scale  and%
\begin{equation}
 STr \left( M^{n}\right) \equiv \sum_{i}\left( -1\right) ^{2j_{i}}\left(
2j_{i}+1\right) m_{i}^{n},
\end{equation}%
is the supertrace, written in terms of the the spin of the
different particles $j_{i}$ and the field-dependent mass
eigenvalues $m_{i}$.

The form of (\ref{Coleman}) gives a field theory interpretation to
the scalar potential found in Section \ref{sec4.1}. Let us try and
match the 1-loop expression with the potential (\ref{Coleman})
interpreting the various terms in the Coleman-Weinberg potential
as different terms in the $\epsilon$ expansion in (\ref{expansion
of V}). We first notice that in any spontaneously broken
supergravity theory, $STr\left( M^{0}\right) =0$, as the number of
bosonic and fermionic degrees of freedom must be equal. The
leading term in (\ref{Coleman}) is therefore null.

We recall that due to the extended no-scale structure the
coefficient of the $\mc{O}(\epsilon)$ term in (\ref{expansion of V}) is also
vanishing. Our comparison should therefore involve the leading
non-zero terms in both cases. In the following paragraphs, we will
re-analyse the three examples studied in Section \ref{sec3} and
show how we always get a matching. This gives a nice
physical understanding of this cancellation at leading order in
$\delta V^{KK}_{(g_{s}),1-loop}$ which is due just to
supersymmetry: the cancellation must take place if the resulting
1-loop potential is to match onto the Coleman-Weinberg form.
Supersymmetry causes the vanishing of the first term in
(\ref{Coleman}) and we notice, for each example, that the second
term in (\ref{Coleman}) scales as the $\mc{O}(\epsilon^2)$ term in (\ref{expansion of
V}), therefore, in order to match the two results, the $\mc{O}(\epsilon)$ term
in (\ref{expansion of V}) also has to be zero. This is, in fact,
what the extended no-scale structure guarantees.

We note here that both, with the use of the supergravity expression for the Coleman-Weinberg
formula, and for the earlier discussions of section 3, supersymmetry has played a crucial role.
In the Coleman-Weinberg formula, the presence of low-energy supersymmetry is used to evaluate the supertraces
and to relate these to the gravitino mass. In the discussion of kinetic terms, the fact that the
corrections are written as corrections to the K\"ahler metric automatically implies that the structure of low-energy
supersymmetry is respected.

\subsubsection{Case 1: \textit{N=1 }$T^{6}/(%
%TCIMACRO{\U{2124} }%
%BeginExpansion
\mathbb{Z}
%EndExpansion
_{2}\times
%TCIMACRO{\U{2124} }%
%BeginExpansion
\mathbb{Z}
%EndExpansion
_{2})$} \label{sec4.2.1}

The case of the \textit{N=1} toroidal orientifold background was
 studied in Section \ref{sec3.1.1} and \ref{sec3.2.1}. We here treat all three
moduli on equal footing, reducing the volume form (\ref{torus}) to
the one-modulus case
\begin{equation}
\mathcal{V}=\tau ^{3/2} = \left(\frac{T+\bar{T}}{2}\right)^{3/2}.
\end{equation}
We therefore take
\begin{equation}
\left\langle \tau _{1}\right\rangle \simeq \left\langle \tau
_{2}\right\rangle \simeq \left\langle \tau _{3}\right\rangle .
\end{equation}%
We write out very explicitly the correction to the scalar
potential due to the correction to the K\"ahler potential as
computed in \cite{bhk}. We focus only on the K\"ahler moduli
dependence. The tree level K\"ahler potential is
$$
K = -3 \ln (T + \bar{T})
$$
and the loop-corrected K\"ahler potential has the form
$$
K = - 3 \ln (T + \bar{T}) + \frac{\epsilon}{(T + \bar{T})}.
$$
The scalar potential is
$$
V = M_P^4 e^K \left( K^{i \bar{j}} \partial_i K \partial_{\bar{j}} K - 3 \right) \vert W \vert^2.
$$
Evaluated, this gives
\bea
V & = & \frac{M_P^4}{(T + \bar{T})^3} \left( 0 + \frac{0 \ti \mc{O}(\epsilon)}{T + \bar{T}} + \frac{\mc{O}(\epsilon^2)}{(T + \bar{T})^2} \right) \nonumber \\
& = & \frac{M_P^4 \epsilon^2}{(T + \bar{T})^5} \sim \frac{M_P^4 \epsilon^2}{\mc{V}^{10/3}}.
\label{potsimp}
\eea
The cancellation of the $\mc{O}(T + \bar{T})^{-3}$ term in (\ref{potsimp}) is due to the original no-scale structure.
The cancellation of the $\mc{O}(T + \bar{T})^{-4}$ term in (\ref{potsimp}) is due to the extended no-scale structure that is
satisfied by the loop corrected K\"ahler potential, giving a leading contribution at $\mc{O}(T + \bar{T})^{-5}$.
This gives the behaviour of the leading contribution to the scalar potential, which we want to compare with the Coleman-Weinberg
expression.

To compare with (\ref{Coleman}) we recall that in supergravity the
supertrace is proportional to the gravitino mass:
\begin{equation}
STr\left( M^{2}\right) \simeq m_{3/2}^{2}.
\end{equation}%
The dependence of the gravitino mass on the volume is always given by%
\begin{equation}
m_{3/2}^{2}=e^{K}W_{0}^{2}\simeq \frac{1}{\mathcal{V}^{2}}\text{ }%
\Longrightarrow \text{ } STr\left( M^{2}\right) \simeq \frac{1}{\mathcal{V}%
^{2}}.  \label{STr}
\end{equation}%
We must also understand the scaling behaviour of the cut-off
$\Lambda $. $\Lambda$ should be identified with the energy scale
above which the four-dimensional effective field theory breaks
down. This is the compactification scale at which many new KK
states appear, and so is given by
\begin{equation}
\Lambda =m_{KK}\simeq \frac{M_{s}}{R}=\frac{M_{s}}{\tau ^{1/4}}=\frac{1}{%
\tau ^{1/4}}\frac{1}{\sqrt{\mathcal{V}}}M_{P}=\frac{M_{P}}{\mathcal{V}^{2/3}}%
.
\end{equation}%
In units of the Planck mass, (\ref{Coleman}) therefore scales as
\begin{eqnarray}
\label{fqx} \delta V_{1loop} &\simeq &0\cdot \Lambda ^{4}+\Lambda
^{2}STr\left( M^{2}\right) +STr\left( M^{4}\ln \left(
\frac{M^{2}}{\Lambda ^{2}}\right)
\right) \simeq   \notag \\
&\simeq &0\cdot \frac{1}{\mathcal{V}^{8/3}}+\frac{1}{\mathcal{V}^{10/3}}+%
\frac{1}{\mathcal{V}^{4}},
\end{eqnarray}%
in agreement with (\ref{potsimp}).

\subsubsection{Case 2: $%
%TCIMACRO{\U{2102} }%
%BeginExpansion
\mathbb{C}
%EndExpansion
P_{[1,1,1,6,9]}^{4}$} \label{sec4.2.2}

This case, studied in Section \ref{sec3.2.2}, is more involved, as
it includes two K\"{a}hler moduli, the large modulus
 $\tau _{b}\simeq \mathcal{V}^{2/3}$ and the small modulus $\tau _{s}$. The
effective potential gets contributions from loop corrections for
both moduli and in these two cases, (\ref{pp}) takes the
form (the dilaton is considered fixed and its dependence is
reabsorbed in $\mathcal{C}^{KK}_{b}$ and $\mathcal{C}^{KK}_{s}$)

\begin{enumerate}
\item Big modulus%
\begin{eqnarray}
\delta V_{\left( g_{s}\right) ,1-loop}^{KK} &=&\left( 0\cdot \frac{\mathcal{C}%
^{KK}_{b}}{\tau _{b}}+\frac{\alpha _{2,b}\left(
\mathcal{C}^{KK}_{b}\right) ^{2}}{\tau
_{b}^{2}}+\frac{\alpha _{3,b}\left( \mathcal{C}^{KK}_{b}\right) ^{3}}{\tau _{b}^{3}%
}+\mathcal{O}\left( \frac{\partial ^{4}K_{0}}{\partial \tau
_{b}^{4}}\right)
\right) \frac{W_{0}^{2}}{\mathcal{V}^{2}}\text{ }  \notag \\
&\simeq &\left( 0\cdot \frac{\mathcal{C}^{KK}_{b}}{\mathcal{V}^{8/3}}+\frac{%
\alpha _{2,b}\left( \mathcal{C}^{KK}_{b}\right) ^{2}}{\mathcal{V}^{10/3}}+\frac{%
\alpha _{3,b}\left( \mathcal{C}^{KK}_{b}\right)
^{3}}{\mathcal{V}^{4}}\right) W_{0}^{2}.  \label{big1}
\end{eqnarray}

\item Small modulus%
\begin{equation}
\delta V_{\left( g_{s}\right) ,1-loop}^{KK}=\left( 0\cdot
\mathcal{C}^{KK}_{s} \frac{\sqrt{\tau
_{s}}}{\mathcal{V}^{3}}+\frac{\alpha _{2,s}\left( \mathcal{C}_{s}
^{KK}\right) ^{2}}{\mathcal{V}^{3}\sqrt{\tau _{s}}}+\frac{\alpha
_{3,s}\left(\mathcal{C}^{KK}_{s}\right) ^{3}}{\mathcal{V}^{3}\tau _{s}^{3/2}}+\mathcal{O}%
\left( \mathcal{V}^{-2}\frac{\partial ^{4}K_{0}}{\partial \tau
_{s}^{4}}\right) \right) W_{0}^{2}.  \label{small1}
\end{equation}
\end{enumerate}

In the Coleman-Weinberg potential, the supertrace has the same
scaling $\sim \mc{V}^{-2}$ as in (\ref{STr}), but there now exist
different values of the cut-off $\Lambda$ for the field theories
living on branes wrapping the big and small
4-cycles%

\medskip

\begin{equation}
\left\{
\begin{array}{c}
\Lambda _{b}=m_{KK,b}\simeq \frac{1}{\tau _{b}^{1/4}}\frac{1}{\sqrt{\mathcal{%
V}}}M_{P}=\frac{M_{P}}{\mathcal{V}^{2/3}}, \\
\Lambda _{s}=m_{KK,s}\simeq \frac{1}{\tau _{s}^{1/4}}\frac{1}{\sqrt{\mathcal{%
V}}}M_{P}.%
\end{array}%
\right.   \label{KK mass scale}
\end{equation}

\medskip

%\begin{figure}[ht]
%\linespread{0.2}
%\begin{center}
\FIGURE{\epsfxsize=0.3\hsize \epsfbox{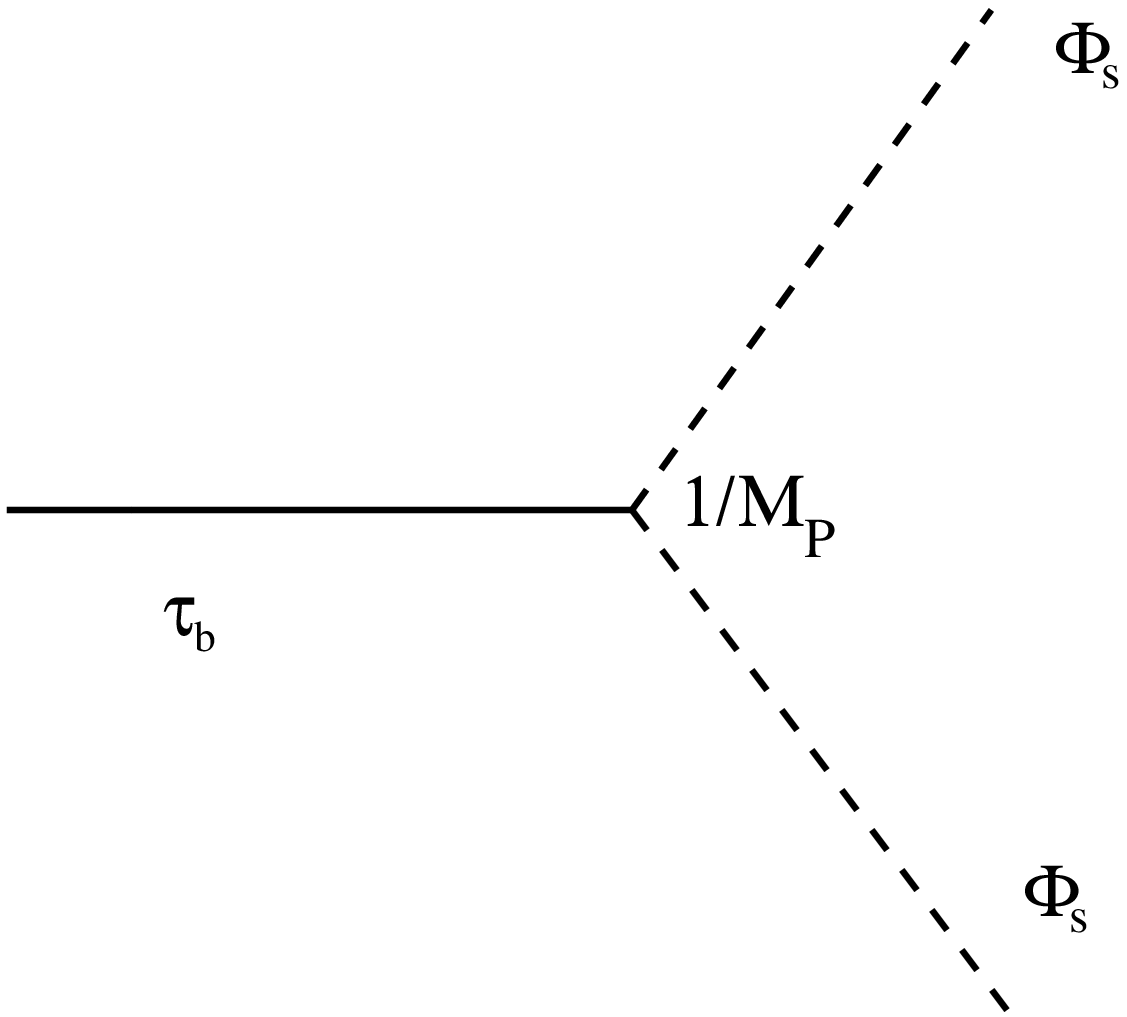}
%\end{center}
\caption{Coupling of the big modulus KK modes to a generic field
  $\Phi_s$ living on the brane wrapping the small 4-cycle. }
\label{diag2}}
%\end{figure}

The existence of two cut-off scales requires some explanation. At
first glance, as $\Lambda _{b}<\Lambda _{s}$ and the KK modes of
the big K\"{a}hler modulus couple to the field theory on the brane
wrapping the small 4-cycle, one might think that there is just one
value of the cut-off $\Lambda$, which is given by $\Lambda_{b}$ =
$m_{KK,b}$. This corresponds to the mass scale of the lowest
Kaluza-Klein mode present in the theory. For a field theory living
on a brane wrapping the large cycle, this represent the mass scale
of Kaluza-Klein replicas of the gauge bosons and matter fields of
the theory. However, we do not think this is the correct
interpretation for a field theory living on the small cycle. The
bulk Kaluza-Klein modes are indeed lighter than those associated
with the small cycle itself.

\medskip

%\begin{figure}[ht]
%\linespread{0.2}
%\begin{center}
\FIGURE{\epsfxsize=0.3\hsize \epsfbox{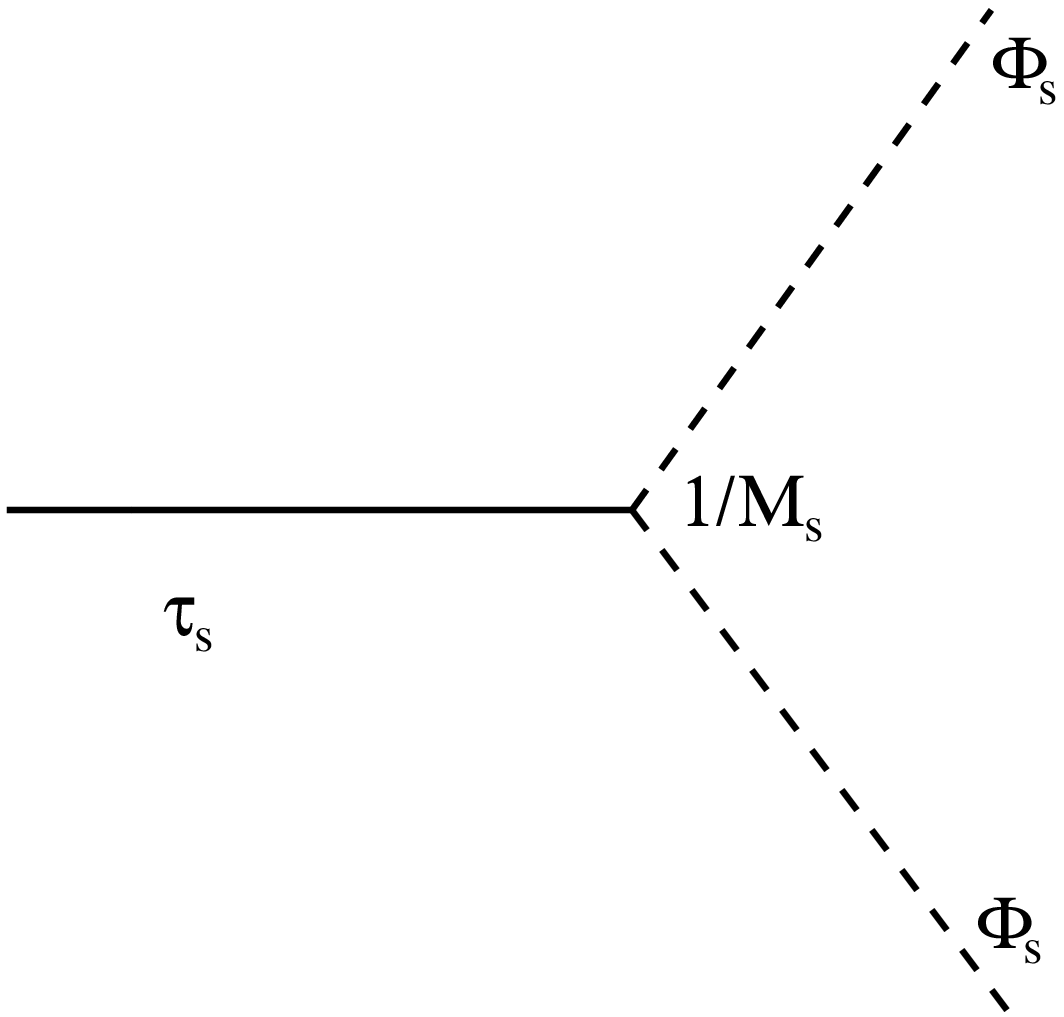}
%\end{center}
\caption{Coupling of the small modulus KK modes to a generic field
  $\Phi_s$ living on the brane wrapping the small 4-cycle.  }
\label{setup}}
%\end{figure}

However it is also the case that the bulk modes couple extremely
weakly to this field theory compared to the local modes. The bulk
modes only couple gravitationally to this field theory, whereas
the local modes couple at the string scale \cite{conlonquev}. In
the case that the volume is extremely large, this difference is
significant. For a field theory on the small cycle, the cutoff
should be the scale at which KK replicas of the quarks and gluons
appear, rather than the scale at which new very weakly coupled
bulk modes are present. As the local modes are far more strongly
coupled, it is these modes that determine the scale of the UV
cutoff. This is illustrated in Figure 2 and 3.\footnote{Notice
that the cut-off dependence of the $STr(M^2)$ term could
potentially be dangerous for the stability of the magnitude of
soft terms computed for this model in references \cite{soft}. With
our analysis here it is easy to see that the contribution of this
term to the scalar potential and then to the structure of soft
breaking terms is suppressed by inverse powers of the volume and
is therefore harmless.}

\bigskip

We now move on to make the
matching
of (\ref{big1}) and (\ref{small1}) with the Coleman-Weinberg potential (\ref%
{Coleman}). For the big modulus, we find
\begin{eqnarray}
\delta V_{1loop} &\simeq &0\cdot \Lambda _{b}^{4}+\Lambda
_{b}^{2}STr\left(
M^{2}\right) +STr\left( M^{4}\ln \left( \frac{M^{2}}{\Lambda _{b}^{2}}%
\right) \right) \simeq   \notag \\
&\simeq &0\cdot \frac{1}{\mathcal{V}^{8/3}}+\frac{1}{\mathcal{V}^{10/3}}+%
\frac{1}{\mathcal{V}^{4}},
\end{eqnarray}%
which yields again a scaling matching that of (\ref{big1}).
For the small modulus we obtain, proceeding as in the previous
case
\begin{eqnarray}
\delta V_{1loop} &\simeq &0\cdot \Lambda _{s}^{4}+\Lambda
_{s}^{2}STr\left(
M^{2}\right) +STr\left( M^{4}\ln \left( \frac{M^{2}}{\Lambda _{s}^{2}}%
\right) \right) \simeq   \notag \\
&\simeq &0\cdot \frac{1}{\tau _{s}}\frac{1}{\mathcal{V}^{2}}+\frac{1}{\sqrt{%
\tau _{s}}}\frac{1}{\mathcal{V}^{3}}+\frac{1}{\mathcal{V}^{4}},
\label{jklm}
\end{eqnarray}%
where we have a matching only of the second term of
(\ref{jklm}) with the second term of (\ref{small1}). This is
indeed the term which we expect to match, given that is the first
non-vanishing leading contribution to the effective scalar
potential at 1-loop. There is no reason the first terms need to
match as they have vanishing coefficients.

As an aside, we finally note that
the third term in (\ref{small1}) can also match with the
Coleman-Weinberg effective potential, although we should not try
to match this with the third term in (\ref{Coleman}) but with a
subleading term in the expansion of the second term in
(\ref{Coleman}). This is due to the fact that we do not have full
control on the expression for the Kaluza-Klein scale (\ref{KK mass
scale}). In the presence of fluxes, this is more reasonably given
by (for example see the discussion in appendix D of \cite{bhp})
\begin{eqnarray}
\Lambda _{s} &=&m_{KK,s}\simeq \frac{1}{\tau _{s}^{1/4}}\frac{M_{P}}{\sqrt{%
\mathcal{V}}}\left( 1+\frac{1}{\tau _{s}}+...\right)
=\frac{1}{\tau
_{s}^{1/4}}\frac{M_{P}}{\sqrt{\mathcal{V}}}+\frac{1}{\tau _{s}^{5/4}}\frac{%
M_{P}}{\sqrt{\mathcal{V}}}+...  \notag \\
&\Longrightarrow &\Lambda _{s}^{2}\simeq \frac{1}{\tau
_{s}^{1/2}}\frac{M_{P}^{2}%
}{\mathcal{V}}+\frac{2}{\tau
_{s}^{3/2}}\frac{M_{P}^{2}}{\mathcal{V}}.
\end{eqnarray}%
This, in turn, produces%
\begin{equation}
\Lambda _{s}^{2}STr\left( M^{2}\right) \simeq \frac{1}{\tau _{s}^{1/2}}\frac{%
1}{\mathcal{V}^{3}}+\frac{2}{\tau
_{s}^{3/2}}\frac{1}{\mathcal{V}^{3}}. \label{bnbn}
\end{equation}%
In this case the second term in (\ref{bnbn}) reproduces the
scaling behaviour of the third term in (\ref{small1}).

\subsubsection{Case 3: $%
%TCIMACRO{\U{2102} }%
%BeginExpansion
\mathbb{C}
%EndExpansion
P_{[1,1,2,2,6]}^{4}$}

In Section \ref{sec3.2.3} we have seen that there are two regimes
where the case of the \textit{K3} Fibration with two K\"{a}hler
moduli can be studied. When the VEVs of the two moduli are of the
same order of
magnitude, they can be treated on equal footing and the volume form %
(\ref{vol11226}) reduces to the classical one parameter example
which, as we have just seen in Section \ref{sec4.2.1}, gives also
the scaling behaviour of the toroidal orientifold case. We do not
need therefore to repeat the same analysis and we automatically
know that the scaling of our general result for the effective
scalar potential at 1-loop matches exactly the Coleman-Weinberg
formula also in this case.

The second situation when $\tau_{2} \gg \tau_{1}$ is more
interesting. The relations (\ref{tay}) tell us that the large
volume limit $\tau_{2} \gg \tau_{1}$ is equivalent to $t_{1} \gg
t_{2}$ and thus they reduce to
\begin{equation}
\tau _{1}=t_{2}^{2}\text{, \ \ \ \ \ }\tau _{2} \simeq
2t_{2}t_{1}, \qquad \mc{V} \simeq \half \sqrt{\tau_1} \tau_2
\simeq t_1 t_2^{2}. \label{stay}
\end{equation}
The KK scale of the compactification is then set by the large
2-cycle $t_1$, \be m_{KK} \sim \frac{M_s}{\sqrt{t_1}} \sim
\frac{M_P}{t_1 t_2}, \ee while in the large volume limit the
gravitino mass is \be m_{3/2} \sim \frac{M_P}{\mc{V}} \sim
\frac{M_P}{t_1 t_2^2}. \ee The bulk KK scale is therefore
comparable to that of the gravitino mass, and it is not clear that
this limit can be described in the language of four-dimensional
supergravity. Let us nonetheless explore the consequences of using
the same analysis as in the previous sections. The evaluation of
(\ref{expansion of V}) gives (reabsorbing the VEV of the dilaton in
$\mathcal{C}^{KK}_{1}$ and $\mathcal{C}^{KK}_{2}$)

\begin{enumerate}
\item Small modulus $\tau_{1}$%
\begin{eqnarray}
\delta V_{\left( g_{s}\right) ,1 loop}^{KK}&\simeq&\left( 0\cdot
\frac{\mathcal{C}^{KK}_{1}}{\tau_{1}\mathcal{V}^{2}}+\frac{\alpha
_{2,1}\left( \mathcal{C} ^{KK}_{1}\right)
^{2}}{\tau_{1}^{2}\mathcal{V}^{2}}+\frac{\alpha _{3,1}\left(
\mathcal{C}^{KK}_{1}\right) ^{3}}{\tau_{1}^{3}\mathcal{V}^{2}}
\right) W_{0}^{2}.  \label{SMALL}
\end{eqnarray}

\item Big modulus $\tau_{2}$%
\begin{equation}
\delta V_{\left( g_{s}\right) ,1 loop}^{KK} \simeq \left( 0\cdot
\mc{C}^{KK}_{2}\frac{\sqrt{\tau_{1}}}{\mathcal{V}^{3}}+ \alpha
_{2,2}\left( \mathcal{C}^{KK}_{2}\right)
^{2}\frac{\tau_{1}}{\mathcal{V}^{4}}+ \alpha _{3,2}\left(
\mathcal{C}^{KK}_{2}\right)
^{3}\frac{\tau_{1}^{3/2}}{\mathcal{V}^{5}}\right) W_{0}^{2}.
\label{BIG}
\end{equation}
\end{enumerate}
Let us now derive the two different values of the cut-off
$\Lambda$ for the field theories living on branes wrapping the big
and small 4-cycles. We realise that the Kaluza-Klein radii for the
two field theories on $\tau_{1}$ and $\tau_{2}$ are given by
\begin{equation}
\left\{
\begin{array}{c}
R _{1}\simeq \sqrt{t_{2}}, \\
R_{2}\simeq \sqrt{t_{1}},
\end{array}%
\right.   \label{KK radii}
\end{equation}
and consequently
\begin{equation}
\left\{
\begin{array}{c}
\Lambda _{1}=m_{KK,1}\simeq \frac{M_{s}}{\sqrt{t_{2}}} \simeq
\frac{1}{\tau_{1}^{1/4}\sqrt{\mathcal{V}}} M_{P},\\
\Lambda _{2}=m_{KK,2}\simeq \frac{M_{s}}{\sqrt{t_{1}}} \simeq
\frac{\sqrt{\tau_{1}}}{\mathcal{V}} M_{P}.
\end{array}%
\right.   \label{KKMassScales}
\end{equation}
We note that $m_{KK,2}$ coincides with the scale of the lightest
KK modes $m_{KK}$. If we try to match the result (\ref{SMALL}) for
the small cycle with the corresponding Coleman-Weinberg potential
for the field theory on $\tau_{1}$
\begin{eqnarray}
\delta V_{1loop} &\simeq &0\cdot \Lambda _{1}^{4}+\Lambda _{1}^{2}
\hbox{STr}\left(
M^{2}\right) + \hbox{STr}\left( M^{4}\ln \left( \frac{M^{2}}{\Lambda _{1}^{2}}%
\right) \right) \simeq   \notag \\
&\simeq &0\cdot \frac{1}{\tau_{1}^2\mathcal{V}^{2}}+\frac{1}{\sqrt{\tau_{1}}\mathcal{V}^{3}}+%
\frac{1}{\mathcal{V}^{4}},
\end{eqnarray}%
we do not find any agreement. This is not surprising since
effective field theory arguments only make sense when
\begin{equation}
\delta V_{\left( g_{s}\right) ,1 loop}^{KK} \ll m_{KK}^{4},
\label{condition}
\end{equation}
but this condition is not satisfied in our case. In fact, using
the mass of the lowest KK mode present in the theory, we have
\begin{equation}
m_{KK}^4 = m_{KK,2}^{4}\simeq\ \frac{\tau_{1}^2}{\mathcal{V}^{4}}
\ll \frac{1}{\tau_{1}^{2}\mathcal{V}^{2}}\simeq\delta V_{\left(
g_{s}\right) ,1 loop}^{KK}.  \label{mkk24}
\end{equation}
Energy densities couple universally through gravity, and so this
implies an excitation of Kaluza-Klein modes, taking us
 beyond the regime of validity of effective field theory.
Thus in this limit the use of the four-dimensional supergravity
action with loop corrections to compute the effective potential
does not seem trustworthy, as it gives an energy density much
larger than $m_{KK}^4$.

For the field theory on the large cycle $\tau_{2}$ the
Coleman-Weinberg potential gives
\begin{eqnarray}
\delta V_{1loop} &\simeq &0\cdot \Lambda _{2}^{4}+\Lambda
_{2}^{2}STr\left( M^{2}\right) +STr\left( M^{4}\ln \left(
\frac{M^{2}}{\Lambda _{2}^{2}}
\right) \right) \simeq   \notag \\
&\simeq &0\cdot
\frac{\tau_{1}^2}{\mathcal{V}^{4}}+\frac{\tau_{1}}{\mathcal{V}^{4}}+
\frac{1}{\mathcal{V}^{4}}. \label{fff}
\end{eqnarray}
In this case the energy density given by the loop corrections
(\ref{BIG}) is (marginally) less than $m_{KK}^4 \simeq\
\tau_{1}^2\mathcal{V}^{-4}$, being smaller by a factor of
$\tau_1$. Equation (\ref{fff}) then matches the result (\ref{BIG})
at leading order.

Again, we also note as an aside that if we expand the KK scale as in
in Section \ref{sec4.2.2}, then we obtain
\begin{eqnarray}
\Lambda _{2} &=&m_{KK,2}\simeq
\frac{\sqrt{\tau_{1}}}{\mathcal{V}}\left( 1+\frac{1}{\tau
_{2}}+...\right)M_{P} \simeq
\left(\frac{\sqrt{\tau_{1}}}{\mathcal{V}}+\frac{\tau_{1}}{\mathcal{V}^{2}}\right)M_{P}
\notag \\
&\Longrightarrow &\Lambda _{2}^{2}\simeq
\left(\frac{\tau_{1}}{\mathcal{V}^{2}}+\frac{\tau_{1}^{3/2}}{\mathcal{V}^{3}}\right)M_{P}^{2}.
\end{eqnarray}
This, in turn, produces
\begin{equation}
\Lambda _{2}^{2} STr\left( M^{2}\right) \simeq
\frac{\tau_{1}}{\mathcal{V}^{4}}+\frac{\tau
_{1}^{3/2}}{\mathcal{V}^{5}}. \label{bnbnyu}
\end{equation}
In this case the second term in (\ref{bnbnyu}) also reproduces the
scaling behaviour of the third term in (\ref{BIG}).

\section{Conclusions}

The purpose of this article has been to study, as far as possible,
the form of loop corrections to the K\"ahler potential for general Calabi-Yau
compactifications and their effect on the scalar potential. The aim has been to
extract the parametric dependence on the moduli that control the loop expansion.
We have contributed to put the proposed form of leading order
string loop corrections on firmer grounds in the sense that they
agree with the low-energy effective action behaviour. In
particular, it is reassuring that the Coleman-Weinberg formula for
the scalar potential fits well with that arising from the BHP conjecture for the corrections to
the K\"ahler potential.
Furthermore, the non-contribution of the leading order string loop
correction is no longer an accident but it is just a manifestation
of the underlying supersymmetry with equal number of bosons and
fermions, despite being spontaneously broken.

These results are important for K\"{a}hler moduli stabilisation.
  In particular, even though the string loop corrections to the
  K\"{a}hler potential
are subdominant with respect to the leading order $\alpha'$
contribution, they can be more important than non-perturbative
superpotential corrections to stabilise non blow-up moduli. The
general picture is that all corrections - $\alpha'$, loop and
non-perturbative - play a r\^{o}le in a generic Calabi-Yau
compactification. We will discuss these matters in more detail in
a forthcoming companion article \cite{ccq2}.

\section{Acknowledgements}
We thank S. AbdusSalam, M. Berg, C. Burgess, F. Denef, A.
Hebecker, C. Kounnas, L. McAllister and K. Suruliz for useful
conversations. We also thank M. Berg, M. Haack and E. Pajer for
interesting comments on the first version of this article.
MC is partially funded by St John's College, EPSRC
and CET. JC is funded by Trinity College, Cambridge. FQ is
partially funded by STFC and a Royal Society Wolfson merit award.

\appendix
\section{Survey of Moduli Stabilisation Mechanisms}
\label{appendix}

We have seen that the no-scale structure of the scalar potential
will be broken by several contributions which will lead to the
following general
form%
\begin{equation}
V=V_{np}+V_{(\alpha'
)}+V_{(g_{s})}^{KK}+V_{(g_{s})}^{W}+V_{local}+V_{D},
\label{general}
\end{equation}%
where $V_{np}$ and $V_{(\alpha ' )}$ are given by (\ref{scalar}),
and $V_{(g_{s})}^{KK}$ and $V_{(g_{s})}^{W}$ are the perturbative
contributions from the string loop corrections (\ref{UUU}) and
(\ref{UUUU}). $V_{local}$ is the potential generated by extra
local sources and $V_{D}$ is the usual D-term scalar potential for
\textit{N=1}\ supergravity

\begin{equation}
V_{(D)}=\frac{1}{2}\left( \left( \hbox{Re}f\right) ^{-1}\right)
^{\alpha \beta }D_{\alpha }D_{\beta },\text{ \ \ \ }D_{\alpha
}=\left[ K_{i}+\frac{W_{i}}{W}\right] \left( T_{\alpha }\right)
_{ij}\varphi _{j}. \label{D scalar}
\end{equation}%
We now review moduli stabilisation mechanisms proposed in the
literature in order to illustrate the importance of having a
deeper understanding of the string loop corrections. From the
expression (\ref{scalar}) we realise that
\begin{equation}
V_{np}\sim e^{K}\left( W_{np}^{2}+W_{0}W_{np}\right) ,\text{ \ \ \
\ \ \ \ \ \ \ }V_{p}\sim e^{K}W_{0}^{2}K_{p},  \label{scaling}
\end{equation}%
where in general we have%
\begin{equation}
V_{p}=V_{(\alpha ' )}+V_{(g_{s})}^{KK}+V_{(g_{s})}^{W},
\end{equation}%
for the full perturbative contributions to the scalar potential.
Let us explore the possible scenarios which emerge by varying
$W_0$. As stressed in Section \ref{sec2.2}, we can trust the use
of solely the leading perturbative corrections to the
scalar potential only when the overall volume is stabilised at large values%
\text{ }$\mathcal{V}\gg 1$. The first systematic study of the
strength of perturbative corrections
to the scalar potential was in \cite{bb}. Neglecting $%
V_{(g_{s})}^{KK}$, $V_{(g_{s})}^{W}$, $V_{local}$ and $V_{(D)}$,
\cite{bb} studied the behaviour of the minima of the scalar
potential when one varies $\left\vert W_{0}\right\vert $. Their
results are summarized in the following table:

\bigskip

\begin{tabular}{|c|c|c|}
\hline 1) $\left\vert W_{0}\right\vert \sim \left\vert
W_{np}\right\vert \ll 1$ &
2) $\left\vert W_{np}\right\vert <\left\vert W_{0}\right\vert <1$ & 3) $%
\left\vert W_{np}\right\vert \ll \left\vert W_{0}\right\vert \simeq \mathcal{%
O}(1)$ \\ \hline $\left\vert V_{(\alpha ' )}\right\vert \ll
\left\vert V_{np}\right\vert $ & $\left\vert V_{np}\right\vert
\simeq \left\vert V_{(\alpha ' )}\right\vert $ & $\left\vert
V_{np}\right\vert \ll \left\vert V_{(\alpha ' )}\right\vert $ \\
\hline
\end{tabular}

\bigskip

\begin{enumerate}
\item $\left\vert W_{0}\right\vert \sim \left\vert W_{np}\right\vert \ll 1$
\ $\Longrightarrow $ $\left\vert V_{(\alpha ' )}\right\vert
/\left\vert V_{np}\right\vert \sim $\ $\left\vert \delta
K_{(\alpha ' )}\right\vert \sim 1/\mathcal{V}\ll 1$ \
$\Longleftrightarrow $ \ $\left\vert V_{(\alpha ' )}\right\vert
\ll \left\vert V_{np}\right\vert $

This case is the well-known KKLT scenario \cite{kklt}. All moduli
are stabilised by non-perturbative corrections at an AdS
supersymmetric minimum with $D_{T}W=0$. A shortcoming of this
model is that $W_{0}$ must be tuned very small in order to
stabilise at large volume and neglect $\alpha '$ or other
perturbative corrections. KKLT gave
the following fit for the one-parameter case:%
\begin{equation}
W_{0}=-10^{-4},\text{ \ }A=1,\text{ \ }a\simeq 2\pi /60\text{ \ \ }%
\Longrightarrow \text{ \ \ }\left\langle \tau \right\rangle \simeq
113\Longleftrightarrow V\simeq 1.2\cdot 10^{2}.
\end{equation}
In addition to $\vert W_0 \vert \ll 1$, a large rank gauge group
(as in $SU(60)$ above) is also necessary to get $a\tau \gg 1$.
This is a bit inelegant but a lower rank of the gauge group would
imply a much worse fine tuning of $W_{0}$. The authors also
proposed a mechanism to uplift the solution to dS, by adding a
positive potential generated by the tension of $\overline{D3}$
branes. This represents an explicit breaking within 4D
supergravity. Remaining within a supersymmetric effective theory,
\cite{bkq} proposed using D-term uplifting to keep manifest
supersymmetry whereas \cite{SS} instead proposed F-term uplifting
using metastable supersymmetry breaking vacua. Also \cite{cnp}
pointed out that the KKLT procedure in two steps (first the
minimisation of $S$
and $%
U_{\alpha }$ at tree level and then $T_{i}$ fixed
non-perturbatively) can miss important contributions such as a dS
minimum without the need to add any up-lifting term.

We finally notice that this mechanism also relies on the
assumption that $W_{np}$ depends explicitly on each K\"{a}hler
modulus. In the fluxless case, this assumption is very strong as
only arithmetic genus 1 cycles \cite{witten} would get stringy
instanton contributions and $D7$ brane deformation moduli would
remain unfixed. The presence of the corresponding extra fermionic
zero modes can prevent gaugino condensation and in general could
also destroy instanton contributions for non-rigid arithmetic
genus 1 cycles. However by turning on fluxes, the D7 moduli should
be frozen and the arithmetic genus 1 condition can be relaxed.
Therefore it is possible that also non-rigid cycles admit
nonperturbative effects.

\item $\left\vert W_{np}\right\vert <\left\vert W_{0}\right\vert
<1\Longrightarrow $ $\left\vert V_{(\alpha ' )}\right\vert
/\left\vert V_{np}\right\vert \sim \left\vert \delta K_{(\alpha '
)}\right\vert
/\left\vert W_{np}\right\vert \left\vert W_{0}\right\vert \sim 1$ \ $%
\Longleftrightarrow $ \ $\left\vert V_{np}\right\vert \simeq
\left\vert V_{(\alpha ' )}\right\vert $

\cite{bb} pointed out that there is an upper bound on the $|W_0|$
in order to find a KKLT minimum $\left\vert W_{0}\right\vert \leq
W_{\max }$. $W_{\max }$ is the value of $\left\vert
W_{0}\right\vert $ for which the leading $\alpha ' $ corrections
start becoming important and compete with the non-perturbative
ones to find a minimum. This minimum will be non-supersymmetric as
we can infer from looking at (\ref{scalar}) which
implies that $V\sim \mathcal{O}(1/\mathcal{V}^{3})$ at the minimum, while $%
-3e^{K}\left\vert W\right\vert ^{2}\sim
\mathcal{O}(1/\mathcal{V}^{2})$. Now since the scalar potential is
a continuous function of $\left\vert
W_{0}\right\vert $, increasing $\left\vert W_{0}\right\vert $ from $%
\left\vert W_{0}\right\vert =W_{\max }-\varepsilon $, where we
have an AdS supersymmetric minimum, to $\left\vert
W_{0}\right\vert =W_{\max }+\varepsilon $, will still lead to an
AdS minimum which is now non-supersymmetric. Subsequently, when
$\left\vert W_{0}\right\vert $ is further increased, the $\alpha '
$ corrections become more and more important and the minimum rises
to Minkowski and then de Sitter and finally disappears. The
disappearance corresponds to the $\alpha'$ corrections completely
dominating the non-perturbative ones and the scalar potential is
just given by the last term in (\ref{scalar}) that has clearly a
runaway behaviour without a minimum.

Unfortunately there is no clear example in the literature that
realizes this situation for $\mathcal{V}\gg 1$. In their analysis
\cite{bb} considered the possibility of getting a Minkowski
minimum for the quintic Calabi-Yau $
%TCIMACRO{\U{2102} }%
%BeginExpansion
\mathbb{C}
%EndExpansion
P_{[1,1,1,1,1]}^{4}$ ($\chi =-200$), giving the following fit
\begin{eqnarray}
W_{0} &=&-1.7,\text{ \ }A=1,\text{ \ }a=2\pi /10,\text{ \ }\xi =0.4,\text{\ }%
\hbox{Re}\left( S\right) =1  \notag \\
&\Longrightarrow &\text{ \ \ }\left\langle \tau \right\rangle
\simeq 5\Longleftrightarrow \mathcal{V}\simeq 2.
\end{eqnarray}

We note that this example, in reality, belongs to the third case since $%
\left\vert W_{0}\right\vert \simeq \mathcal{O}(1)$ where we
claimed that no minimum should exist. That is true only for
$\mathcal{V}\gg 1$ but in this case $\mathcal{V}\simeq 2$ and the
higher $\alpha'$ corrections cannot be neglected anymore. Moreover
with $g_{s}\simeq 1$ the string loop expansion is uncontrolled.

\item $\left\vert W_{np}\right\vert \ll \left\vert W_{0}\right\vert \sim
\mathcal{O}(1)$ \ $\Longrightarrow $ \ $\left\vert V_{(\alpha '
)}\right\vert /\left\vert V_{np}\right\vert \sim \left\vert \delta
K_{(\alpha ' )}\right\vert /\left\vert W_{np}\right\vert \gg 1$ \ $%
\Longleftrightarrow $ \ \ $\left\vert V_{(\alpha ' )}\right\vert
\gg \left\vert V_{np}\right\vert $

This is the more natural situation when $\left\vert
W_{0}\right\vert \sim \mathcal{O}(1)$. In this case if we ignore
the non-perturbative corrections and keep only the $\alpha ' $
ones no minimum is present.
However there are still $V_{(g_{s})}^{KK}$, $%
V_{(g_{s})}^{W}$, $V_{local}$ and $V_{D}$. Thus, let us see two
possible scenarios

\begin{enumerate}
\item $V_{np}$ neglected, $V_{(\alpha ' )}+ V_{local}$ considered

Bobkov \cite{bobkov} considered F-theory compactifications on an
elliptically-fibered Calabi-Yau four-fold $X$ with a warped
Calabi-Yau three-fold $M$ that admits a conifold singularity at
the base of the fibration. Following the procedure proposed by
Saltman and Silverstein \cite{SS2} for flux compactifications on
products of Riemann surfaces, he added $n_{D7}$ additional pairs
of $D7/ \overline{D7}$-branes and $n_{7}$ extra pairs of $(p,q)$
$7/\overline{7}$-branes wrapped around the 4-cycles in $M$ placed
at the loci where the fiber $T^{2}$ degenerates. These extra local
sources generate positive tension and an anomalous negative
$D3$-brane tension contribution to $V_{local}$ which, in units of
$\left(
\alpha ' \right) ^{3}$, reads%
\begin{eqnarray}
V &=&-\chi \left( 2\pi \right) ^{13}N_{flux}^{2}\left( \frac{g_{s}^{4}}{%
\mathcal{V}_{s}^{3}}\right) -N_{7} \left( \frac{%
g_{s}^{3}}{\mathcal{V}_{s}^{2}}\right)   \notag \\
&&+n_{7}\left( \frac{g_{s}^{2}}{\mathcal{V}_{s}^{4/3}}\right)
+n_{D7}\left( \frac{g_{s}^{3}}{\mathcal{V}_{s}^{4/3}}\right) ,
\end{eqnarray}
where ${\mathcal{V}_{s}}$ is the string frame volume and
$N_{7}=\left( n_{D7}^{3}+n_{7}^{3}\right)$ is an effective
parameter given in terms of triple intersections of branes. By
varying the various parameters, this is argued to give a
discretuum of large-volume non-supersymmetric AdS, Minkowski and
metastable dS vacua for Calabi-Yau three-folds with $h_{1,1}=1$
(this implies $\chi <0$). The fit proposed is for the dS solution:
\begin{eqnarray}
\left\vert W_{0}\right\vert  &\simeq &\left( 2\pi \right) ^{2}N_{flux}>1,%
\text{ }\chi =-4,\text{ }N_{flux}=3,\text{ }n_{7}=1,\text{
}n_{D7}=73,
\notag \\
\text{ }g_{s} &\simeq &5\cdot 10^{-3}\text{ \ }\Longrightarrow \text{ \ \ }%
\mathcal{V}\simeq 3\cdot 10^{4}.
\end{eqnarray}

The integer parameters are tuned to obtain a pretty small $g_{s}$
so that the effect of string loop corrections can be safely
neglected. In this scenario, in which supersymmetry is broken at
the Kaluza-Klein scale, the stabilisation procedure depends on
local issues, while we would prefer to have a more general
framework where we could maintain global control.

\item $V_{np}$ neglected, $V_{(\alpha ' )}+V_{(g_{s})}^{KK}+$ $%
V_{(g_{s})}^{W}$ considered

Berg, Haack and K\"{o}rs \cite{bhk}, following their exact
calculation of
the loop corrections for the \textit{N=1} toroidal orientifold $T^{6}/(%
%TCIMACRO{\U{2124} }%
%BeginExpansion
\mathbb{Z}
%EndExpansion
_{2}\times
%TCIMACRO{\U{2124} }%
%BeginExpansion
\mathbb{Z}
%EndExpansion
_{2})$ analyzed if these corrections could compete with the
$\alpha '$ ones to generate a minimum for $V$. By treating the
three toroidal K\"{a}hler moduli in $T^{6}=T^{2}\times T^{2}\times
T^{2}$ on an equal footing they reduce the problem to
a 1-dimensional one. They neglect $%
V_{g_{s}}^{KK}$ as it is suppressed by higher powers of the
dilaton and compare just $V_{\alpha ' }$ and $V_{g_{s}}^{W}$. The
schematic form of the scalar potential is \be V \sim \frac{\xi
\vert W_0 \vert^2}{\mc{V}^3} + \frac{\delta}{\mc{V}^{10/3}}. \ee
It turns out that $\delta > 0$, and so as $\xi \sim - \chi$, they
need a positive Euler number $\chi >0$ in order to find a minimum,
while the $T^6/(\mbb{Z}_2 \ti \mbb{Z}_2)$ toroidal example has a
negative Euler number. They instead consider the \textit{N=1}
toroidal orientifold $T^{6}/\mbb{Z}_6'$ that satisfies the
condition $\chi >0$. A non-supersymmetric AdS minimum is now
present but as the loop corrections are naturally subleading with
respect to the $\alpha ' $\ ones, they must
fine tune the complex structure moduli to get large volume. They find%
\begin{eqnarray}
\left\vert W_{0}\right\vert  &\sim &\mathcal{O}(1),\text{ \ }\hbox{Re}%
(U)\simeq 650,\text{\ }\hbox{Re}\left( S\right) =10  \notag \\
\text{ \ } &\Longrightarrow &\text{ \ \ }\left\langle \tau
\right\rangle \simeq 10^{2}\Longleftrightarrow \mathcal{V}\simeq
10^{3}.
\end{eqnarray}
The fine-tuning comes from assuming the complex structure moduli
are stabilised at large values. A similar scenario has been
studied also by von Gersdorff and Hebecker \cite{hg}. In addition,
Parameswaran and Westphal \cite{para} studied the possibility to
have a consistent D-term uplifting to de Sitter in this scenario.
\end{enumerate}

%\bigskip

\item We have assumed above that when $\left\vert
W_{0}\right\vert \sim \mathcal{O}(1)$ perturbative corrections
always dominate non-perturbative ones, which can therefore be
neglected. But is this naturally always the case? In order to
answer this question, let us now consider scenarios in which
 $V_{np}$ and $V_{(\alpha
' )}$ compete while $\left\vert W_{0}\right\vert \sim \mathcal{O}%
(1)$.
\begin{enumerate}
\item[(a)] \bigskip $V_{(g_{s})}^{KK}+$ $V_{(g_{s})}^{W}$ neglected, $%
V_{np}+V_{(\alpha ' )}$ considered $\ \Longrightarrow $ \ large
volume

This situation was studied by Westphal \cite{Westphal} following
the work of Balasubramanian and Berglund, finding a dS minimum at
large volume for the quintic. However this result extends to other
Calabi-Yau three-folds with just one K\"{a}hler modulus. He
presents the following fit
\begin{eqnarray}
W_{0} &=&-1.7,\text{ \ }A=1,\text{ \ }a=2\pi /100,\text{ \ }\xi =79.8,\text{%
\ }\hbox{Re}\left( S\right) =1  \notag \\
\text{ \ } &\Longrightarrow &\text{ \ \ }\left\langle \tau
\right\rangle \simeq 52\Longleftrightarrow \mathcal{V}\simeq 376.
\end{eqnarray}

The non-perturbative corrections are rendered important by using a
large-rank gauge group $SU(100)$ for gaugino condensation. This is
not fine-tuned but is contrived. The loop corrections, which may
be important, are not considered here.

\bigskip

\item[(b)] $V_{(g_{s})}^{KK}+$ $V_{(g_{s})}^{W}$ neglected, $%
V_{np}+V_{(\alpha ' )}$ considered $\ \Longrightarrow $ \
exponentially large volume
\end{enumerate}

This situation is appealing since it provides a positive answer to
our basic question. Balasubramanian, Berglund and two of the
present authors \cite{bbcq} developed these scenarios which now
go under the name of Large Volume Models, which is a bit
misleading as large volume is always necessary to trust a
solution. They should be more correctly called LARGE Volume Models
because the volume is exponentially large.
 In this framework, both
non-perturbative and $\alpha'$ corrections compete naturally to
get a non-supersymmetric AdS minimum of the scalar potential at
exponentially large volume. This is possible by considering more than one K%
\"{a}hler modulus and taking a well-defined large volume limit.
For one modulus models, the work of \cite{bb} and \cite{Westphal}
shows that with the rank of the gauge group $SU(N)$ in the natural
range $N\simeq 1\div 10$, it is impossible to have a minimum.

However, if we have more generally $h_{1,1}>1$, this turns out to
be possible. The simplest example of such models is for the
hypersurface
$%
%TCIMACRO{\U{2102} }%
%BeginExpansion
\mathbb{C}
%EndExpansion
P_{[1,1,1,6,9]}^{4}$.  The overall volume in terms
of 2-cycle volumes is given by%
\begin{equation}
\mathcal{V}=\frac{1}{6}\left(
3t_{1}^{2}t_{5}+18t_{1}t_{5}^{2}+36t_{5}^{3}\right),
\end{equation}%
and the 4-cycle volumes take the form%
\begin{equation}
\tau _{4}=\frac{t_{1}^{2}}{2}\text{, \ \ \ \ \ }\tau
_{5}=\frac{\left( t_{1}+6t_{5}\right) ^{2}}{2},  \label{LVS}
\end{equation}%
for which it is straightforward to see that%
\begin{equation}
\mathcal{V}=\frac{1}{9\sqrt{2}}\left( \tau _{5}^{3/2}-\tau
_{4}^{3/2}\right) .  \label{volume}
\end{equation}%
The reason why $\tau_{4}$ and $\tau_{5}$ are considered instead of
$\tau_{1}$ and $\tau_{5}$ as outlined in Section \ref{sec2.1}, is
that these are the only 4-cycles which get instanton contributions
to $W$ when fluxes are turned off \cite{ddf}. As we will describe
in our companion paper \cite{ccq2}, to get LARGE Volume Models, we
require that $W_{np}$ depends only on blow-up modes which resolve
point-like singularities, as $\tau_{4}$ in this case. Such cycles
are always rigid cycles and thus naturally admit nonperturbative
effects.
If we now take the large volume limit in the following way%
\begin{equation}
\left\{
\begin{array}{c}
\tau _{4}\text{ small}, \\
\text{\ }\tau _{5}\gg 1,
\end{array}
\right.   \label{assumption}
\end{equation}
the scalar potential looks like%
\begin{equation}
V=V_{np}+V_{(\alpha ' )}\sim \frac{\lambda \sqrt{\tau _{4}}%
e^{-2a_{4}\tau _{4}}}{\mathcal{V}}-\frac{\mu \tau _{4}e^{-a_{4}\tau _{4}}}{%
\mathcal{V}^{2}}+\frac{\nu }{\mathcal{V}^{3}},\text{ \ }\lambda
,\text{ }\mu ,\text{ }\nu \text{ constants}  \label{minimumy}
\end{equation}
with a non-supersymmetric AdS minimum located at%
\begin{equation}
\tau _{4}\sim \left( 4\xi \right) ^{2/3}\text{ \ \ and \ \
}\mathcal{V}\sim \frac{\xi ^{1/3}\left\vert W_{0}\right\vert
}{a_{4}A_{4}}e^{a_{4}\tau _{4}}. \label{minimum}
\end{equation}

The result that we have found confirms the consistency of our
initial
assumption (\ref{assumption}) in taking the large volume limit. Inserting in %
(\ref{minimum}) the correct parameter dependence and with the
following
natural choice of parameters, we find%
\begin{eqnarray}
W_{0} &=&1,\text{ \ }A_{4}=1,\text{ \ }a_{4}=2\pi /7,\text{ \ }\xi =1.31,%
\text{\ }\hbox{Re}\left( S\right) =10  \notag \\
\text{ \ } &\Longrightarrow &\text{ \ \ }\left\langle \tau
_{4}\right\rangle \simeq 41\Longleftrightarrow \mathcal{V}\simeq
3.75\cdot 10^{15}. \label{exp}
\end{eqnarray}
Therefore $\tau _{4}$ is stabilised small whereas $\tau _{5}\gg
1$, and
the volume can be approximated as%
\begin{equation}
\mathcal{V}\sim \tau _{5}^{3/2},
\end{equation}%
and%
\begin{equation}
\tau _{4}\sim t_{1}^{2},\ \ \ \ \ \tau _{5}\sim t_{5}^{2}.
\end{equation}

Looking at (\ref{minimum}) we can realise why in this case we are
able to make
$V_{np}$ and $V_{(\alpha ' )}$ compete naturally. In fact, in general $%
V_{(\alpha ' )}\sim 1/\mathcal{V}^{3}$ and $V_{np}\sim
e^{-a_{4}\tau
_{4}}/\mathcal{V}^{2}$, but (\ref{minimum}) implies $V_{np}\sim 1/\mathcal{V}%
^{3}\sim V_{(\alpha ' )}$. The non-perturbative corrections in the
big modulus $\tau _{5}$ will be, as usual, subleading. An
attractive feature of these models is that they provide a method
of generating hierarchies.
In fact the result (\ref{exp}), for $%
M_{P}\sim 2.4$ $10^{18}$ GeV, produces an intermediate string scale%
\begin{equation}
M_{s}\simeq \frac{M_{P}}{\sqrt{\mathcal{V}}}\sim 10^{11} \hbox{
GeV}, \label{string}
\end{equation}
and this can naturally give rise to the weak scale through
TeV-scale
supersymmetry%
\begin{equation}
M_{soft}\sim m_{3/2}=e^{K/2}\left\vert W\right\vert \sim \frac{M_{P}}{%
\mathcal{V}}\sim 30\hbox{ TeV}.
\end{equation}%

This setup naturally fixes all the moduli while generating
hierarchies. However, it ignores further perturbative corrections
as the $g_{s}$\ ones. It is thus crucial to check if they do not
destroy the picture. Berg, Haack and Pajer applied their guess
(\ref{UUU}) to derive these string loop corrections to the
K\"{a}hler potential. From (\ref{UUU}) it is straightforward to
get \footnote{We note that in this case, as argued by Curio and
Spillner
\cite{curio}, $\delta K_{(g_{s})}^{W}$ is absent, because in $%
%TCIMACRO{\U{2102} }%
%BeginExpansion
\mathbb{C}
%EndExpansion
P_{[1,1,1,6,9]}^{4}$ there is no intersection of the divisors that
give rise
to nonperturbative superpotentials if wrapped by D7 branes.}%
\bea
\delta K_{(g_{s})}^{KK} & \sim & \frac{\mathcal{C}_{4}^{KK}\sqrt{\tau _{4}}}{%
\hbox{Re}\left( S\right)
\mathcal{V}}+\frac{\mathcal{C}_{5}^{KK}\sqrt{\tau
_{5}}}{\hbox{Re}\left( S\right) \mathcal{V}},  \label{AA} \\
\delta K_{(g_{s})}^{W} & \sim & \frac{\mathcal{C}_{4}^{W}}{%
\hbox{Re}\left( S\right){\sqrt{\tau _{4}}
\mathcal{V}}}+\frac{\mathcal{C}_{5}^{W}} {\hbox{Re}\left( S\right)
\sqrt{\tau _{5}} \mathcal{V}}.  \label{AB}
\eea%
 The corrections %
(\ref{AA}) turn out to yield subleading corrections to the scalar
potential of the form
\begin{equation}
V_{(g_{s})}^{KK}\sim \frac{\left( \mathcal{C}_{4}^{KK}\right) ^{2}W_{0}^{2}}{%
\hbox{Re}\left( S\right) ^{2}\mathcal{V}^{3}\sqrt{\tau _{4}}}+O(\mathcal{V}%
^{-10/3}),
\end{equation}%
even if one tries to fine tune the coefficients
$\mathcal{C}_{4}^{KK}$ pretty large, $\mathcal{C}_{4}^{KK}\simeq
20\div 40$. We therefore conclude that the LARGE Volume Scenario
is safe.
\end{enumerate}

This survey of moduli stabilisation mechanisms has shown that a
deeper understanding of string loop corrections to the K\"{a}hler
potential in Calabi-Yau backgrounds is highly desirable. In KKLT
stabilisation, the magnitude of the perturbative corrections is
what determines the regime of validity of the stabilisation
method. In all other methods of stabilisation, perturbative
corrections enter crucially into the stabilisation procedure, and
so not only $\alpha'$ but also $g_{s}$ corrections should be taken
into account.

These loop corrections are neglected in the cases (3a), (4a) and
(4b), but we learnt from the case (3b) that they can change the
vacuum structure of the system studied. However in this situation
a significant amount of fine tuning was needed to make them
compete with the $\alpha'$ corrections to produce a minimum at
large volume. In case (4b), the loop corrections did not
substantially affect the vacuum structure unless they were
fine-tuned large.
 Therefore one would tend to conclude that these string loop
corrections will in general be subdominant and so that it is safe
to neglect them.

While this may be true for models with relatively few moduli, we
will see in \cite{ccq2} that loop corrections can still play a
very important r\^{o}le in moduli stabilisation, in particular
lifting flat directions in LARGE Volume Models. In this case the
fact that they are subdominant will turn out to be a good property
of these corrections since they can lift flat directions without
destroying the minimum already found in the other directions of
the K\"{a}hler moduli space.

\end{document}